\PassOptionsToPackage{dvipsnames}{xcolor}
\documentclass[manuscript,nonacm]{acmart}
\AtBeginDocument{%
  }

\usepackage{hyperref}
\usepackage{listings}
\usepackage{braket}
\usepackage{svg}
\usepackage{graphicx}
\usepackage{subcaption}
\usepackage{amsmath}

\DeclareMathOperator*{\argmin}{arg\,min}

\newtheorem{theorem}{Theorem}[section]

\newtheorem{definition}[theorem]{Definition}

\newtheorem{proposition}[theorem]{Proposition}

\newtheorem{lemma}[theorem]{Lemma}

\usepackage{multirow}
\usepackage{array}
\newcommand{\code}{\texttt}

\begin{document}

\title{A Qubit-Efficient Hybrid Quantum Encoding Mechanism for Quantum Machine Learning}

\author{Hevish Cowlessur}
\email{hcowlessur@student.unimelb.edu.au}
\orcid{0009-0006-4999-4597}
\affiliation{%
  \institution{The University of Melbourne}
  \city{Melbourne}
  \state{Victoria}
  \country{Australia}
}
\affiliation{%
  \institution{CSIRO's Data61}
  \city{Sydney}
  \country{Australia}
}

\author{Tansu Alpcan}
\orcid{0000-0002-7434-3239}
\affiliation{%
  \institution{The University of Melbourne}
  \city{Melbourne}
  \country{Australia}}
\email{tansu.alpcan@unimelb.edu.au}

\author{Chandra Thapa}
\orcid{0000-0002-3855-3378}
\affiliation{%
  \institution{CSIRO's Data61}
  \city{Sydney}
  \country{Australia}
}
\email{chandra.thapa@data61.csiro.au}

\author{Seyit Camtepe}
\orcid{0000-0001-6353-8359}
\affiliation{%
  \institution{CSIRO's Data61}
  \city{Sydney}
  \country{Australia}
}
\email{seyit.camtepe@data61.csiro.au}

\author{Neel Kanth Kundu}
\orcid{0000-0002-6439-4024}
\affiliation{%
  \institution{Indian Institute of Technology Delhi}
  \city{New Delhi}
  \country{India}
}
\email{neelkanth@iitd.ac.in}

\renewcommand{\shortauthors}{H. Cowlessur et al.}

\begin{abstract}
Efficiently embedding high-dimensional datasets onto noisy and low-qubit quantum systems is a significant barrier to practical Quantum Machine Learning (QML). 
Approaches such as quantum autoencoders can be constrained by current hardware capabilities and may exhibit vulnerabilities to reconstruction attacks due to their invertibility. 
We propose Quantum Principal Geodesic Analysis (qPGA), a novel, non-invertible method for dimensionality reduction and qubit-efficient encoding. Executed classically, qPGA leverages Riemannian geometry to project data onto the unit Hilbert sphere, generating outputs inherently suitable for quantum amplitude encoding. This technique preserves the neighborhood structure of high-dimensional datasets within a compact latent space, significantly reducing qubit requirements for amplitude encoding. 
We derive theoretical bounds quantifying qubit requirements for effective encoding onto noisy systems. Empirical results on MNIST, Fashion-MNIST, and CIFAR-10 show that qPGA preserves local structure more effectively than both quantum and hybrid autoencoders. Additionally, we demonstrate that qPGA enhances resistance to reconstruction attacks due to its non-invertible nature. 
In downstream QML classification tasks, qPGA can achieve over 99\% accuracy and F1-score on MNIST and Fashion-MNIST, outperforming quantum-dependent baselines. Initial tests on real hardware and noisy simulators confirm its potential for noise-resilient performance, offering a scalable solution for advancing QML applications.
\end{abstract}

\keywords{qPGA, Data pre-processing,
  Quantum Machine Learning, Quantum Amplitude Encoding, Principal Geodesic Analysis}
\maketitle

\section{Introduction}
Real-world datasets are getting increasingly high-dimensional, sparse, and complex \cite{10.1109/TPAMI.2013.50}. Across fields such as finance, healthcare, image processing, or natural language processing, this poses significant challenges for classical computational algorithms \cite{doi:10.1126/science.aaa8415}. Quantum computing is often advocated as a promising solution to these challenges \cite{schuld2015introduction}. 
However, leveraging the capabilities of quantum computing critically depends on our ability to effectively encode high-dimensional classical data onto quantum systems \cite{ciliberto2018quantum}. 
This challenge is particularly pressing in the Noisy Intermediate-Scale Quantum (NISQ) era, characterized by constrained qubit availability and elevated noise levels \cite{Preskill2018quantumcomputingin}. Even for the future of fault-tolerant quantum computing, the investigation of data pre-processing and feature extraction algorithms tailored to quantum applications, such as Quantum Machine Learning (QML), while optimizing quantum resources, remains critical for the scalability and efficiency of quantum computing. Further complicating matters, QML models such as variational quantum circuits (VQCs) often exhibit barren plateaus--regions of vanishing gradients--especially when the encoding stage involves a large number of qubits, which typically leads to a higher number of trainable parameters in the ansatz \cite{qi2023barren}. This makes it particularly important to explore encoding techniques that not only embed high-dimensional data onto small qubit systems but also reduce circuit complexity \cite{cerezo2021variational}.

While classical pre-processing techniques such as Principal Component Analysis (PCA) and Linear Discriminant Analysis (LDA) have been used in QML pipelines for feature extraction and dimensionality reduction, they are fundamentally designed for Euclidean spaces~\cite{mancilla_preprocessing_2022}. Hence, these techniques are often misaligned with the structure of quantum state spaces, which have non-Euclidean geometry, and may not directly preserve meaningful structure during quantum encoding.  In the quantum domain, existing methods for feature extraction via dimensionality reduction for encoding of high-dimensional datasets onto small-qubit quantum systems have been proposed, but many of these approaches have limited empirical validation on practical benchmark datasets \cite{lloyd_quantum_2014, yu_quantum_2019, cong_quantum_2016}, or rely on quantum hardware components like quantum RAM or oracles \cite{tabi_ieee, sakhnenko_hybrid_2022, chao-yang_design_2006, kerenidis_classification_2020, romero_quantum_2017}, which, in the NISQ-era, are limited by qubit number and noise for practical deployment. 
Moreover, while quantum-dependent autoencoders, such as Quantum Autoencoders (QAEs) \cite{romero_quantum_2017}, and Hybrid Quantum Autoencoders (HAEs) \cite{sakhnenko_hybrid_2022, tabi_h}, have shown effectiveness in dimensionality-reduction, they often face security risks due to their potential inversion to reconstruct data \cite{wang2022data}. 
Given these limitations, there is a need for innovative approaches that can effectively bridge the gap between high-dimensional datasets and the constraints of quantum systems, enabling non-invertible pre-processing and qubit-efficient encoding--one that can effectively embed high-dimensional data, maintaining its key characteristics, onto small qubit systems.

In this work, we present Quantum Principal Geodesic Analysis (qPGA)--a novel, non-invertible feature extraction and hybrid encoding method that is designed to be resistant to reconstruction attacks. 
Crucially, the qPGA algorithm itself is executed entirely using classical computational resources. Its ``quantum'' aspect lies in its design principles being inspired by, and its output being tailored for, subsequent quantum amplitude encoding. While the foundational geometric and dimensionality reduction machinery of qPGA is adapted from classical Principal Geodesic Analysis (PGA) and its kernelized versions (kPGA), the innovation here lies in identifying the Unit Hilbert Sphere (UHS) as the native geometric space for real-valued classical vectors when normalized for quantum amplitude encoding. qPGA's specific value proposition in the quantum context is its ability to respect the non-Euclidean geometry of these amplitude-encoded quantum state vectors, a contrast to classical PCA, which operates in Euclidean space.
\textit{To the best of our knowledge, qPGA is the first to leverage the geometry of unit-normed classical data vectors representing amplitude-encoded quantum states (which we refer to as \textit{amplitude vectors}), for embedding high-dimensional datasets onto small-qubit systems}. 
qPGA studies the geometric structure of these vectors--which lie on a unit Hilbert sphere (UHS)--using statistical tools from Riemannian geometry \cite{sommer2010manifold, geomstats}. Our approach is a quantum-inspired adaptation of the kernel Principal Geodesic Analysis (kPGA) algorithm, which, in the classical domain, has been proposed for hyperspherical statistical analyses in reproducing kernel Hilbert space (RKHS) \cite{calders_kernel_2014}. By operating in this geometric framework, qPGA enables meaningful dimensionality reduction tailored to the structure of amplitude-encoded data. 
Specifically, qPGA preserves both the local neighborhood structure and global variance of the dataset in its low-dimensional latent space representations, which cannot be inverted using standard geometric methods to recreate the original high-dimensional dataset. Crucially, our proposed methodology allows data pre-processing, enabling informative and structure-preserving encoding onto small-qubit quantum systems for downstream QML tasks through classical methods--effectively decoupling feature extraction from quantum hardware.

\begin{figure}[ht]
    \centering
    \includegraphics[width=\linewidth]{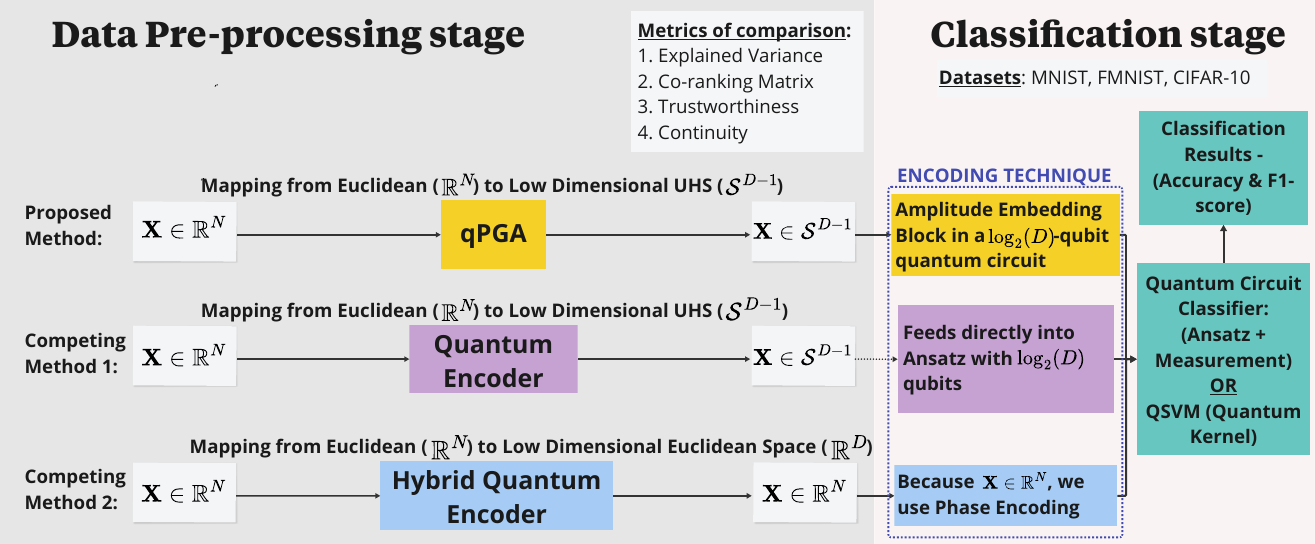}
    \caption{High-level workflow of this paper. We compare our proposed qPGA method against Quantum Encoder and Hybrid Quantum Encoder as 2 competing methods for encoding high-dimensional classical data onto small-qubit systems. The three methods are evaluated on four metrics and employed in two classification pipelines using QSVM and Quantum Circuit Classifier.}
    \Description{Workflow of paper}
    \label{fig:initial_flow}
\end{figure}

In this paper, we first formalize the proposed qPGA algorithm. Then, we derive theoretical bounds for the number of qubits required to efficiently encode high-dimensional datasets onto noisy small-qubit systems. These bounds inform the practical qubit requirements for maintaining the rate of qubit errors within acceptable thresholds, enabling the design of scalable and noise-resilient QML pipelines with high-dimensional real-valued datasets. 
To assess its effectiveness in the quantum domain, we benchmark qPGA against two established quantum encoder methods: the Quantum Autoencoder and the Hybrid Quantum Autoencoder.
Our experiments demonstrate that qPGA consistently outperforms these baselines in both the data pre-processing stage with respect to preserving local data structure and in two end-to-end classification pipelines using distinct QML models when suitable kernel functions are chosen, underscoring its versatility. 
Additionally, we show that qPGA exhibits better resistance in an adversarial reconstruction attack setup. Finally, real hardware experiments provide preliminary signs of qPGA's practical potential for enabling small-qubit systems to perform effectively in near-term QML applications. We summarize the overall workflow of this paper in Fig. \ref{fig:initial_flow}.

\noindent
Our key contributions in this paper are as follows: 
\begin{enumerate}
    \item \textbf{qPGA algorithm}: We propose the qPGA algorithm as a feature extraction via dimension reduction methodology for allowing qubit-efficient amplitude encoding of real-valued datasets. Our method reduces the qubit requirements for encoding high-dimensional datasets by preserving their local structure in their latent space better than the evaluated quantum encoder-based methods. We used four metrics to demonstrate this. 
    \item \textbf{Theoretical qubit bounds}: We provide theoretical analyses of the link between the number of qubits required to capture a target dataset variance, and the presence of qubit error or noise, derived under an assumption of independent qubit errors. These bounds offer a valuable initial framework for resource estimation under idealized noise conditions, paving the way for future refinements incorporating more complex noise phenomena.
    \item \textbf{Inherent robustness}: We empirically demonstrate that our proposed qPGA algorithm exhibits greater robustness to adversarial reconstruction attacks on latent data compared to quantum-dependent encoders due to its lossy compression nature and reduced invertibility when using standard geometric inverse operations. 
    \item \textbf{qPGA for QML classification}: We demonstrate two use cases of the proposed qPGA algorithm for QML classification tasks on three widely used benchmark datasets, using two distinct quantum classifiers—the Quantum Support Vector Classifier and Quantum Circuit Classifier. These use cases highlight qPGA's effectiveness across different classification paradigms, achieving high accuracy without requiring deep or wide quantum circuits. Further experiments on actual devices and noisy simulators provide initial evidence of qPGA's practical potential in downstream QML tasks using current-generation quantum devices.
\end{enumerate}
We structure the rest of this paper as follows. 
Section \ref{literature} reviews the relevant literature. 
In Section \ref{problem_statement}, we scope and describe the research problem addressed in this work, and Section \ref{sec:kpga_for_quantum} introduces our qPGA algorithm as a hybrid methodology for encoding high-dimensional datasets onto small qubit systems. In Section \ref{theory}, we establish theoretical bounds on the number of qubits required for effective data embedding in noisy quantum systems using the qPGA algorithm. Section \ref{sec:exp_results} details our experimental evaluation of qPGA against competing methods in the data pre-processing stage for QML. Next, section \ref{sec:reconstruction} examines the invertibility of qPGA in an adversarial reconstruction attack scenario. Section \ref{sec:classification} presents two end-to-end QML classification pipelines incorporating qPGA. In Section \ref{sec:actual_device_results}, we report results for tests carried out on actual quantum hardware and noisy simulators. Finally, Section \ref{sec:conclusion} summarizes our overall findings and conclusions. 

\section{Related Work \label{literature}}
In this section, we review literature relevant to this work, focusing primarily on existing feature extraction and dimensionality-reduction techniques in the quantum domain. We also describe the kernel Principal Geodesic Analysis (kPGA) algorithm as studied in the classical domain, which serves as a foundation for the qPGA algorithm proposed in this paper.
\subsection{Quantum Algorithms for Dimensionality-reduction \label{lit_rev1}}
Quantum algorithms offer promising approaches for feature extraction via dimensionality reduction during the data pre-processing stages, with potential advantages over classical methods in terms of computational efficiency. Some key techniques from previous works are Quantum Principal Component Analysis (qPCA), Quantum Linear Discriminant Analysis (QLDA), and Quantum Slow Feature Analysis (QSFA). 
Lloyd et al. \cite{lloyd_quantum_2014} introduced qPCA with the potential for exponential acceleration over classical PCA for certain input states.
Similarly, QLDA, \cite{cong_quantum_2016, yu_quantum_2023}, is a supervised technique with theoretical proof of speed-ups that aims to find a lower-dimensional subspace that maximizes class separability. Moreover, the QSFA algorithm, implemented for reducing the dimensions of classical data suitable for QML pipelines, assumes access to quantum memory~\cite{kerenidis_classification_2020}.

However, the practical implementation of these algorithms is constrained by the current limitations of quantum hardware, such as restricted qubit counts and susceptibility to noise. Many of these approaches rely on assumptions--such as access to quantum memory or noise-free input states--that are not yet feasible on near-term devices. As a result, their immediate applicability to data pre-processing in QML remains limited. Moreover, efficient resource utilization under these constraints is still an active area of research, highlighting the need for dimensionality-reduction techniques that are not only theoretically sound but also compatible with today’s noisy quantum systems. 
Thus, to avoid reliance on limited quantum hardware during the pre-processing stage, we devise qPGA as a hybrid method to encode high-dimensional classical data as low-dimensional amplitude vectors of quantum states in their computational bases. This approach minimizes dependence on quantum hardware during the pre-processing stage by performing dimensionality reduction classically, enabling high-dimensional classical data to be efficiently mapped onto quantum states with fewer qubits for downstream QML tasks. By handling dimensionality reduction on classical hardware, qPGA avoids the need for resource-intensive quantum subroutines (e.g., qPCA and QSFA), making the pre-processing pipeline more compatible with NISQ devices. The resulting low-dimensional representations can then be amplitude-encoded on small-qubit quantum systems with reduced error and resource requirements.
\subsection{Autoencoders in the Quantum Domain}
Romero et al. introduced the quantum autoencoder (QAE), which consists of a quantum neural network model designed to encode and compress information stored in quantum states \cite{romero_quantum_2017}. It is inspired by classical autoencoders but operates on quantum data and circuits. Similar to the classical analog,  QAE consists of 3 layers: the input/encoder layer, the bottleneck layer, and the output/decoder layer. At the bottleneck layer, the compressed latent space representation consists of the quantum states represented by a subset of the initial number of qubits.  
In this work, we use the QAE architecture introduced in \cite{romero_quantum_2017} to compress high-dimensional data into their low-dimensional representations in the quantum domain, as a competing method benchmarking our proposed qPGA algorithm. While QAE offers an effective mechanism for feature extraction and dimensionality reduction, it faces challenges when applied to high-dimensional datasets. These challenges stem from the limited number of qubits and elevated noise levels in current quantum devices, making it difficult to effectively encode high-dimensional benchmark datasets. Even for the fault-tolerant generation of quantum devices, embedding high-dimensional classical data onto QAE would require large numbers of qubits and hence large numbers of trainable parameters. Such large circuits become harder to train due to the vanishing gradient problem \cite{qi2023barren}. 

Often, to overcome the limitations due to the high dimensionality of data, classical neural network layers are used for feature extraction, followed by a quantum circuit in the bottleneck layer. This is the Hybrid Quantum Autoencoder (HAE), where the quantum part of the encoder improves the data represented in the latent space. HAE has been widely studied in literature \cite{sakhnenko_hybrid_2022, srikumar_clustering_2022, tabi_ieee}. The work in \cite{sakhnenko_hybrid_2022} utilized phase embeddings for the quantum circuit of the hybrid encoder. However, since each phase embedding gate can only encode a single data point per gate, this makes the HAE qubit-hungry if we have a relatively high-dimensional latent space. Moreover, the limitations due to the current generation of quantum devices also apply to HAE, similar to QAE. 

Hence, a quantum computing-independent method for pre-processing classical data for their efficient embeddings onto small-qubit systems is needed to bypass the limitations involved in employing a quantum-dependent pre-processing method. In this work, we propose qPGA as such a method. It is important to note that while qPGA is classical, its design is explicitly tailored for preparing data for quantum amplitude encoding. This distinguishes it from general-purpose classical dimensionality reduction techniques that do not inherently consider the geometric constraints of quantum state spaces, particularly the Unit Hilbert Sphere relevant to amplitude-encoded states. 
\subsection{Kernel Principal Geodesic Analysis (kPGA)}
In the classical domain, Awate et al. proposed the kPGA algorithm as an adaptation of (Euclidean) Kernel Principal Component Analysis (kPCA) \cite{scholkopf1998nonlinear}, suitable for hyperspherical statistical analysis in kernel feature space \cite{calders_kernel_2014}. This method combines concepts from kernel methods and Riemannian geometry to perform dimension reduction and statistical analyses on nonlinear data. kPGA builds upon the foundations of Principal Geodesic Analysis (PGA) \cite{pga} and kernel methods, which map data into a high-dimensional feature space to enable linear analysis of nonlinear data. This approach allows for the separation of complex data structures that may not be linearly separable in the original space. The method operates on the Riemannian manifold of the Hilbert sphere in the reproducing kernel Hilbert space (RKHS). This geometric perspective allows for the analysis of data that lies on nonlinear Riemannian manifolds such as the unit Hilbert sphere (UHS) rather than linear Euclidean spaces.
 
Thus, in this work, motivated by the foundational principles of kPGA, we propose qPGA as a novel approach for feature extraction for the quantum domain. 
The novelty of qPGA lies not in the inventions of PGA or kPGA themselves, but in its specific adaptation and application to the pre-processing of classical data destined for quantum amplitude encoding. By recognizing that real-valued classical vectors, when normalized for amplitude encoding, naturally inhabit the surface of a UHS, qPGA leverages the geometric tools from Riemannian geometry to find a lower-dimensional representation on a lower-dimensional UHS that is optimized for qubit efficiency.
Specifically, our qPGA algorithm allows the encoding of high-dimensional datasets as latent space vectors that represent low-dimensional amplitude-encoded quantum states for small-qubit systems. Our approach ensures that these amplitude vectors retain a high proportion of the original dataset's variance while preserving its intrinsic neighborhood structure, facilitating efficient embedding onto small-qubit systems.

\section{Problem Statement\label{problem_statement}}
In this section, we identify and define our research problem by breaking it down into specific research questions to address the stated research problem in a systematic way. 
We state our research problem as follows: \\
\textit{How can we embed high-dimensional benchmark datasets onto small-qubit systems to minimize quantum resource consumption and the effect of noise, while maintaining essential dataset characteristics in their latent-space representations, thereby enabling qubit-efficient and noise-aware quantum encoding for downstream classification tasks?}

Dimensionality reduction algorithms for feature extraction for the quantum domain are necessary to keep the size of qubit systems\footnote{Qubit systems here can refer to quantum circuits that are used in QML algorithms, which require encoding of classical data as quantum states for processing onto a quantum computer.} small, making efficient use of quantum resources, to keep computational complexity low, and ensure the trainability of QML models. Moreover, feature extraction is needed because the encoding of high-dimensional data on quantum hardware is limited by the number of qubits and circuit size due to noise and errors, specifically on current-generation devices. 

Existing methods for feature extraction via dimensionality reduction for the quantum domain, such as quantum autoencoders and hybrid quantum-classical autoencoders, rely heavily on quantum hardware for their practical implementation. As a result, they still require encoding high-dimensional datasets onto quantum circuits--a qubit-intensive process. These approaches can hence be constrained due to the limitations of current devices and would prevent efficient use of critical quantum resources that could otherwise be used for tasks such as classification and optimization. Hence, to overcome these limitations, we propose an alternative embedding approach that avoids early reliance on quantum hardware. We focus on classically pre-processed amplitude vectors and define our research scope through the following sub-questions:
\begin{enumerate}
    \item How can we effectively characterize and study the geometric structure of the space occupied by high-dimensional amplitude vectors to enable efficient feature extraction for encoding onto small-qubit systems?
    \item What theoretical and experimental frameworks can be established to robustly evaluate and validate our proposed embedding methodology? 
\end{enumerate}
By investigating these questions, our goal is to develop a robust, resource-efficient, and scalable solution for data encoding that maintains key dataset characteristics and enhances performance for QML applications, both in and beyond the NISQ era. 
\section{\label{sec:kpga_for_quantum}Quantum Principal Geodesic Analysis (qPGA) for Quantum Amplitude Encoding}
In this section, we describe the qPGA algorithm proposed in this paper as the main data pre-processing methodology for high-dimensional classical datasets to encode their latent space representations as amplitude vectors of quantum states of a small-qubit system. 
It is crucial to reiterate that the qPGA algorithm itself is executed entirely using classical computational resources; its ``quantum'' aspect lies in its design principles being inspired by and tailored for quantum amplitude encoding. This approach can be considered a ``Hybrid Quantum Encoding Strategy,'' where qPGA pre-processes data for efficient quantum encoding. 

In quantum computing, amplitude encoding is a technique used to encode classical datasets into the amplitudes of quantum states, which we describe next. 
While there are other quantum encoding techniques, such as phase and basis encoding, in this work, we focus on the amplitude encoding technique, which is better suited for developing our algorithm.

Given a real-valued $N$-dimensional dataset $\mathbf{X} = \{ \mathbf{x}_i \in \mathbb{R}^N \}^M_{i=1}$, where $M$ is the number of data points in the dataset, each data point $\mathbf{x}_i$ is normalized and encoded into the normalized amplitudes of a quantum state $| \psi_i \rangle$ in its computational basis states, $| j \rangle$, such that:
\begin{equation}
| \psi_i \rangle = \sum_{j=1}^{N} \mathbf{x}_{ij} | j \rangle, \quad \text{where all } \mathbf{x}_{ij} \in \mathbb{R} \text{, satisfying} \sum_{j=1}^{N} |\mathbf{x}_{ij}|^2 = 1.
\label{amplitudeEmbedding}
\end{equation}
In this work, since we focus on encoding real-valued datasets--which is also the main focus of most works by the QML community--it is sufficient to assume that all $\mathbf{x}_{ij} \in \mathbb{R}$, rather than $\mathbf{x}_{ij} \in \mathbb{C}$, which is generally the case in the amplitude encoding technique, to develop our algorithm. In amplitude encoding, a real-valued vector of dimension $N$ is mapped to the amplitudes of a quantum state in a $2^n$-dimensional Hilbert space, with $n$ qubits satisfying $2^n \geq N$. If $N$ is not a power of two, zero-padding is applied to extend the vector to the nearest power of two. As such, representing a dataset with $N$ features requires $\lceil\log_2N\rceil$ qubits.
The ceiling function ensures that we have an integer number of qubits sufficient for encoding amplitude vectors of dimension $N$. 

\textit{However, exact encoding of an $N$-dimensional vector ($N=2^n$) generally requires $\mathcal{O}(2^n)$ gates, which becomes infeasible for large $n$ \cite{PhysRevResearch.4.023136}.} This scaling implies that embedding high-dimensional datasets demands exponentially deep circuits \cite{PhysRevA.102.032420}, rendering amplitude encoding impractical without prior dimensionality reduction. Hence, this motivates the investigation of efficient and scalable feature extraction methods tailored for quantum amplitude encoding.

In this work, we propose a dimensionality reduction method that classically studies the manifold structure of amplitude vectors from high-dimensional real-valued datasets. Specifically, when represented as amplitude vectors, data points reside on the surface of an $N$-dimensional unit Hilbert sphere (UHS), $\mathcal{S}^{N-1}$. Our approach aims to efficiently obtain latent space embeddings that preserve intrinsic geometric properties, including local neighborhood structure and dataset variance, thereby respecting the underlying nonlinear geometry of the data.

Our primary contribution is the development of quantum Principal Geodesic Analysis (qPGA), which extends and adapts kernel Principal Geodesic Analysis (kPGA)~\cite{calders_kernel_2014} for the quantum domain, specifically addressing critical constraints of QML. Unlike existing data pre-processing methods for QML, qPGA explicitly targets the inherent nonlinear geometry of quantum amplitude vectors lying on the unit Hilbert sphere (UHS). By directly preserving the geometric structure intrinsic to quantum states, qPGA provides a fundamentally novel approach to embedding high-dimensional classical data onto resource-limited quantum hardware. Consequently, qPGA not only targets current encoding bottlenecks but also enables quantum models to efficiently capture the essential statistical and geometric characteristics of complex datasets, representing a significant advancement over existing quantum embedding techniques.

\subsection{\label{qPGA_alg}The qPGA Algorithm}
qPGA is designed to reduce the dimensionality of data points $\{\mathbf{x}_i\}_{i=1}^{M}\in \mathbb{R}^N$ lying on the UHS $\mathcal{S}^{N-1}$. By applying principles from Riemannian geometry, qPGA ensures that the intrinsic manifold structure is preserved during the dimensionality-reduction process. This is crucial for maintaining data characteristics, such as local neighborhood structure and dataset variance, when mapping the latent-space representations from the qPGA algorithm to low-qubit quantum circuits as amplitude vectors. Hence, unlike existing data pre-processing techniques, qPGA is a novel methodology that studies \textit{classically} the nonlinear space occupied by amplitude vectors--representing high-dimensional amplitude-encoded quantum states--to preserve data geometry in the latent space for efficient encoding onto small-qubit systems. Thus, we describe qPGA as a qubit-efficient hybrid quantum encoding mechanism. 

Before describing the qPGA algorithm for encoding classical data as the amplitudes of quantum states of small-qubit systems, we define a few concepts and terminologies from Riemannian manifold geometry \cite{geomstats}, reformulated for amplitude vectors lying on a UHS. These form the foundation of our qPGA algorithm for quantum encoding.

\begin{definition}[Logarithmic Map]\label{def1}  The logarithmic map $\text{Log}_\mu$ : $\mathcal{S}^{N-1} \rightarrow T_\mu\mathcal{S}^{N-1}$ at a point $\mu$ in a unit Hilbert sphere $\mathcal{S}^{N-1}$ is a function that maps an amplitude vector $\mathbf{x}_i \in \mathcal{S}^{N-1}$ to a tangent vector in the tangent space $T_\mu\mathcal{S}^{N-1}$ at $\mu$. The logarithmic map of $\mathbf{x}_i$ with respect to $\mu$ can be computed as: 
\begin{equation}
\label{eq:logmap}
\text{Log}_\mu(\mathbf{x}_i) = \frac{\mathbf{x}_i - \langle \mathbf{x}_i, \mu \rangle \mu}{\|\mathbf{x}_i - \langle \mathbf{x}_i, \mu \rangle \mu\|} \arccos(\langle \mathbf{x}_i, \mu \rangle).
\end{equation}
\end{definition}
\begin{definition}[Geodesic Distance]\label{def2}  The geodesic distance $d_g(\mu, \mathbf{x}_i)$ between point $\mu$ and amplitude vector $\mathbf{x}_i$ lying on the unit Hilbert sphere $\mathcal{S}^{N-1}$ is defined as the length of the shortest path (geodesic) connecting $\mu$ and $\mathbf{x}_i$ on the sphere.  The geodesic distance between $\mu$ and $\mathbf{x}_i$ on the unit Hilbert sphere is computed as the norm of the logarithmic map:
\begin{equation}
\label{eq:geodesic_distance}
d_g(\mu, \mathbf{x}_i) = ||\text{Log}_\mu(\mathbf{x}_i)||.
\end{equation}
\end{definition} 
\begin{definition}[Exponential Map]\label{def3}  The exponential map $\text{Exp}_\mu$: $T_\mu\mathcal{S}^{N-1}\rightarrow\mathcal{S}^{N-1}$ at a point $\mu$ in a unit Hilbert sphere $\mathcal{S}^{N-1}$ is a function that projects a tangent vector $\mathbf{v}_i$ in the tangent space $T_\mu\mathcal{S}^{N-1}$ at $\mu$ back to a point, representing the amplitude vector $\mathbf{x}_i$, on the unit Hilbert sphere $\mathcal{S}^{N-1}$. For a tangent vector $\mathbf{v}_i\in T_\mu\mathcal{S}^{N-1}$, the projected amplitude vector $\mathbf{x}_i=$ Exp$_\mu(\mathbf{v}_i) \in \mathcal{S}^{N-1}$ can be computed as: 
\begin{equation}
\label{eq:exp_map}
    \mathbf{x}_i = \text{Exp}_\mu(\mathbf{v}_i) = \cos(\|\mathbf{v}_i\|)\mu + \sin(\|\mathbf{v}_i\|) \frac{\mathbf{v}_i}{\|\mathbf{v}_i\|},
\end{equation}
where $\|\mathbf{v}_i\|$ is the norm of the tangent vector $\mathbf{v}_i$. We note that since, $\text{Exp}_\mu(\mathbf{v}_i)$ maps the tangent vector $\mathbf{v}_i$ to the unit Hilbert sphere, $\mathcal{S}^{N-1}$, it has unit norm, i.e., we can write $\|\text{Exp}_\mu(\mathbf{v}_i)\|_{\mathcal{S}^{N-1}} = 1$.
\end{definition}
\begin{definition}[Sample Frechet Mean]\label{def4} The sample Frechet (or Karcher) mean of a set of $M$ amplitude vectors, $\{\mathbf{x}_i \in \mathcal{S}^{N-1}\}^M_{i=1}$, on the unit Hilbert sphere $\mathcal{S}^{N-1}$ is the point on the sphere that minimizes the sum of squared geodesic distances to all $\mathbf{x}_i$ in the set. Given a set of amplitude vectors $\{\mathbf{x}_1, \mathbf{x}_2,\dots, \mathbf{x}_M\}$ on a unit Hilbert sphere $\mathcal{S}^{N-1}$, the sample Frechet mean $\mu$ is defined as
\begin{equation}
    \label{eq:frechet}
    \mu = \argmin_{\mathbf{y} \in \mathcal{S}^{N-1}} \sum_{i=1}^M d_g^2(\mathbf{y}, \mathbf{x}_i).
\end{equation}
\end{definition}

\begin{figure}[!htpb]
    \centering
    \includegraphics[width=\linewidth]{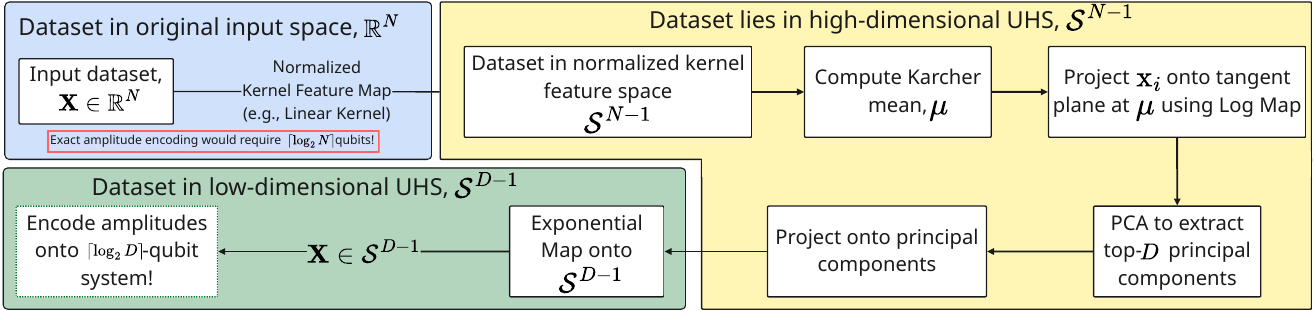} 
    \caption{The qPGA algorithm. We map the input dataset $\mathbf{X}\in\mathbb{R}^N$ onto a high-dimensional UHS, $\mathcal{S}^{N-1}$, perform PCA onto the tangent vectors to extract top-$D$ principal components. We then project the $D$-dimensional tangent vectors onto a low-dimensional UHS, $\mathcal{S}^{D-1}$, comprising D-dimensional amplitude vectors that can now be encoded onto $\lceil\log_2D\rceil$-qubit systems.}
    \Description{qPGA algorithm picture representation}
    \label{fig:qPGA_algorithm}
\end{figure}

Next, using the above definitions and concepts, we describe our qPGA algorithm for encoding real-valued classical datasets, $\mathbf{X}$, as the amplitude vectors of the quantum states of small-qubit systems. Fig. \ref{fig:qPGA_algorithm} gives an illustration of how we map datasets from high-dimensional UHS, $\mathcal{S}^{N-1}$, to low-dimensional UHS, $\mathcal{S}^{D-1}$, for encoding the dataset as the amplitudes of a $\lceil\log_2D\rceil$-qubit system, as per our proposed qPGA algorithm.  
\begin{enumerate}
    \item \textbf{Mapping to a UHS.} We first normalize the dataset $\mathbf{X} = \{\mathbf{x}_i\}^M_{i=1}$, so that each $\mathbf{x}_i$ lies on the UHS, $\mathcal{S}^{N-1}$. We can also apply a normalized kernel feature map $\phi: \mathbb{R}^N \rightarrow \mathcal{S}^{M'-1}$, where $M' > N$ (typically $M'=N$ if no explicit kernel trick to higher dimensions is used, or $M'$ is the dimension of the RKHS if a kernel trick is applied), to map the data to a high-dimensional UHS, $\mathcal{S}^{M'-1}$, ensuring nonlinear relationships are captured.

    \item \textbf{Projection to the tangent space, $T_\mu\mathcal{S}^{N-1}$.} Next, we compute the \textit{Sample Frechet Mean}, $\mu$ as defined in Eq. (\ref{eq:frechet}). We then project the amplitude vectors onto the tangent space of the UHS at $\mu$. Using Eq. (\ref{eq:logmap}), we apply the logarithmic map $Log_\mu: \mathcal{S}^{N-1}\rightarrow T_\mu\mathcal{S}^{N-1}$, to project each amplitude vector $\mathbf{x}_i$ onto the tangent space $T_\mu\mathcal{S}^{N-1}$ at $\mu$ as tangent vectors $Log_\mu(\mathbf{x}_i)$.

    \item \textbf{Dimension Reduction on tangent space vectors.} The tangent space $T_\mu\mathcal{S}^{N-1}$ provides local linearizations of the manifold, such that the logarithmic maps of the amplitude vectors $\{Log_\mu\mathbf{x}_i\}^M_{i=1}$ are now distributed in a linear subspace of $\mathbb{R}^N$ \cite{geomstats}. This allows us to apply techniques in linear manifolds, such as Principal Component Analysis (PCA). By performing PCA on the tangent vectors, $Log_\mu(\mathbf{x}_i)$, we can extract the directions of maximum variance. We first compute the covariance matrix, $\mathbf{C}$ as
\begin{equation}
     \mathbf{C} = \frac{1}{M} \sum_{i=1}^{M} Log_\mu(\mathbf{x}_i) Log_\mu(\mathbf{x}_i)^\top, 
     \label{eq:covariance_matrix}
\end{equation}
and then extract the first $D$-eigenvectors $\mathbf{v}_1, \dots, \mathbf{v}_D$ corresponding to the largest eigenvalues, forming a $D$-dimensional subspace where $D < N$. We next project the tangent vectors $Log_\mu(\mathbf{x}_i)$ onto the $D$-dimensional subspace in the tangent space, capturing maximal variance from the original dataset.

    \item \textbf{Projection onto a low-dimensional UHS, $\mathcal{S}^{D-1}$.} By applying the exponential map $Exp_\mu: T_\mu \mathcal{S}^{D-1} \rightarrow S^{D-1}$ as per Definition \ref{def3}, to the tangent vectors projected onto the $D$-dimensional subspace, we map these vectors onto a $D$-dimensional UHS as $\mathbf{\hat{x}}_i \in \mathcal{S}^{D-1}$. These vectors now specifically represent $D$-dimensional amplitude vectors for a $\lceil\log_2D \rceil$-qubit system. The space occupied by these vectors preserves the geometric structure of the original data points, $\mathbf{x}_i\in\mathbb{R}^N$, in their latent representations, $\mathbf{\hat{x}}_i$.

    \item \textbf{Encoding onto qubit systems.} Each $\mathbf{\hat{x}}_i$ now represents a $D$-dimensional amplitude vector of a $q$-qubit system that is represented by $D = 2^q$ quantum states in their computational bases. Thus, for each $D$-dimensional amplitude vector $\mathbf{\hat{x}}_i \in \mathcal{S}^{D-1}$, the corresponding quantum state $| \psi_i \rangle$ in a qubit system with $q=\lceil\log_2D\rceil$ qubits is given by
   \begin{equation}   
        | \psi_i \rangle = \sum_{j=1}^{D} \mathbf{\hat{x}}_{ij} | j \rangle, \quad \text{where}  \sum_{j=1}^{D} |\mathbf{\hat{x}}_{ij}|^2 = 1.
   \end{equation}
\end{enumerate}

Since only $D$ ($<N$) principal components are required to capture maximal variance from the dataset, discarding the $N-D$ least significant components, we require only 
$\lceil\log_2D\rceil(< \lceil\log_2N\rceil)$ qubits to now encode the dataset into a qubit system that now requires fewer qubits than prior to applying the qPGA algorithm. \textit{Our algorithm, thus, allows us to devise a qubit-efficient encoding mechanism that captures the intrinsic structure of high-dimensional amplitude vectors in its latent space representations.}

So far, we have developed the qPGA algorithm for noise-tolerant qubit systems. However, for current-generation quantum devices, we extend the qPGA algorithm in the next section to account for noise in qubit systems.
\section{Theoretical Bounds for the qPGA Algorithm for Noisy Qubit Systems\label{theory}}

In this section, we provide theoretical bounds on the number of qubits required to efficiently encode the $D$ principal components from our proposed qPGA algorithm, given noisy qubit systems. The performance of our qPGA algorithm during the data pre-processing stages is heavily dependent on the number of principal components, $D$, we extract while performing PCA on the tangent space vectors, as explained in Step (3) above. The larger $D$, the $D$-dimensional subspace consisting of the first $D$-eigenvectors $\mathbf{v}_1, \dots, \mathbf{v}_D$ better represents the high-dimensional data. However, for noisy qubit systems, we need to limit the number of principal components $D$ we can extract using our qPGA algorithm to ensure we are within the system's maximal permissible error bounds. In this section, we provide bounds for the number of qubits $q$ required to represent $D$ principal components from the qPGA algorithm, given a maximum acceptable overall qubit system error (or noise) of $\mathcal{P}_{max}$.

\paragraph{\textbf{Theoretical framework for applying qPGA for noisy qubit systems.}} Firstly, we show in Lemma \ref{lemma1}, the relationship between the number of qubits $q$, and the percentage dataset variance $\beta$, we can capture using $D$ principal components while performing PCA in the high-dimensional tangent space. Then, in Lemma \ref{lemma2}, we define the maximum acceptable overall qubit system noise, $\mathcal{P}_{max}$ and establish an upper bound for the number of qubits, $q$, we can achieve given $\mathcal{P}_{max}$ and the probability of individual qubit error $p$.

\begin{lemma}[Effect of $\beta$ on $q$\label{lemma1}]
Given we want to capture a percentage $\beta$ of the total variance, $\sigma_T^2$, in a dataset of dimension $N$, then the number of qubits, $q$, required to represent the amplitudes of the normalized quantum state vectors, captured by the first $D$ principal components, with eigenvalues $\{\lambda_i\}^D_{i=1}$, is given by
\begin{equation}
q \geq \lceil log_2D \rceil, \text{ where } D = \argmin_k \left\{ \sum_{i=1}^{k} \lambda_i \geq \beta \sigma_T^2\right\}. 
\label{eq:Q}
\end{equation}
\end{lemma}

\noindent
\begin{proof} 
The proof follows from Eq. (\ref{eq:covariance_matrix}). Given the covariance matrix $\mathbf{C}$ is always symmetric and positive semi-definite, we can decompose it using eigenvalue decomposition as:
\begin{equation}
    \mathbf{C} = U \Lambda U^\top,
\end{equation}
\noindent
where \( U \) is the matrix of eigenvectors of $\mathbf{C}$, and $\Lambda=\text{diag}(\lambda_1,\lambda_2,\ldots,\lambda_N)$ and \( \lambda_1 \geq \lambda_2 \geq \ldots \geq \lambda_N \geq 0\). The eigenvalues $\lambda_i$ represent the variance captured by the principal components along the direction of their corresponding eigenvectors in $U$. Then, the total variance $\sigma_T^2$, captured by all principal components, is given by the sum of all eigenvalues: 
\begin{equation}
    \sigma_T^2 = \sum_{i=1}^{N} \lambda_i
\end{equation}
The variance captured by the first \( k \) principal components is given by
 $\sum_{i=1}^{k} \lambda_i$. 
Hence, to capture at least a percentage \( \beta \) of the total variance, we obtain $D \leq 2^q$ as :
\begin{equation}
2^q \geq D = \argmin_{k} \left\{\gamma = \frac{\sum_{i=1}^{k} \lambda_i}{\sigma_T^2}\right\}, \text{ such that } \gamma \geq \beta, 
\label{eq:above}
\end{equation}
where $\gamma$ represents the cumulative explained variance captured by the first $D$ principal components. We select $D$ as the smallest integer for which the inequality in (\ref{eq:above}) holds, leading to a lower bound on the number of qubits required by a $q$-qubit system as stated in Lemma \ref{lemma1}.
\end{proof}

Next, in Lemma \ref{lemma2}, we describe the relationship between qubit error rate, $p$, and the number of qubits, $q$, given a maximal permissible error rate in the system of $\mathcal{P}_{max}$.
\noindent
\begin{lemma}[Effect of qubit error rate $p$ on $q$\label{lemma2}]
Given $\mathcal{P}_{max}$, the maximum acceptable error rate for a $q$-qubit system, where for each qubit, P(error) = $p$, we can bound $q$ above as:
\begin{equation}
q \leq \left\lfloor \frac{\log(1-\mathcal{P}_{max})}{\log(1-p)} \right\rfloor. 
\label{eq:q_pmax}
\end{equation}
\end{lemma}
The floor function is used for rounding down to the nearest integer, ensuring the error constraint is respected.

\noindent
\begin{proof}
The maximum acceptable error rate $\mathcal{P}_{max}$ is equivalent to the probability of at least 1 qubit in a $q$-qubit system being erroneous\footnote{`Erroneous' here encapsulates all the different types of qubit errors, such as dephasing, depolarizing, and gate operation errors.}. Then, assuming $p$ represents the probability of error for each qubit, and that the errors occur separately and independently of each qubit, the probability of maintaining an error-free quantum state is given by $1-\mathcal{P}_{max}=(1-p)^q$. The overall system error, $\mathcal{P}$, is then bounded above by $\mathcal{P}_{max}$ as:
\begin{equation}
    \mathcal{P} \leq \mathcal{P}_{max} = 1-(1-p)^q,
    \label{eq:overall_system_error}
\end{equation}
\noindent
and as the qubit number $q$ grows, the upper-bound $\mathcal{P}_{max}$ also increases. By setting a maximum acceptable error rate $\mathcal{P}_{max}$, we can thus obtain the upper-bound set in Eq. (\ref{eq:q_pmax}), that defines an upper limit on the qubit number $q$ ensuring that the system's error rate remains within acceptable bounds.
\end{proof}
Next, we combine Lemmas \ref{lemma1} and \ref{lemma2} in the proposition below.

\begin{proposition}[Bounds for the efficient encoding of $D$ principal components in noisy qubit systems]\label{proposition}
Let $D$ be the number of principal components obtained from the qPGA algorithm, and $\mathcal{P}_{max}$ be the maximum acceptable error rate for a $q$-qubit system, with P(error)=$p$. Then, for efficiently encoding the $D$ principal components of a dataset within a qubit system, while accounting for both dimension-reduction and qubit noise constraints, the number of qubits $q$ required is determined as: 
\begin{equation}
    \lceil log_2D \rceil \leq q \leq \left\lfloor \frac{\log(1-\mathcal{P}_{max})}{\log(1-p)} \right\rfloor,
\end{equation}
where $D$ is determined as per Eq. (\ref{eq:Q}), representing the number of principal components required to capture a percentage $\beta$ of the total variance $\sigma_T^2$ in the dataset, as established in Lemma \ref{lemma1}.
\end{proposition}
The proof for this proposition follows immediately from the proofs of Lemmas \ref{lemma1} and \ref{lemma2}. This proposition provides theoretical bounds for the quantum system to have a sufficient number of qubits to efficiently encode the principal components of the dataset while also ensuring that the error rate remains within acceptable bounds.

\paragraph{\textbf{Implications of Noise on Qubit Requirements.}} We can conclude from Proposition \ref{proposition} that in noisy quantum systems, the efficiency of encoding and the fidelity of the encoded quantum state are compromised. Increased noise levels impact the qubit requirements for effectively implementing the qPGA algorithm, emphasizing the balance between achieving desired computational precision and maintaining manageable error rates. 
We acknowledge that Lemma 5.2 and Proposition 5.3 rely on the assumption of independent qubit errors. In realistic quantum systems, errors can be correlated due to factors like crosstalk, spatially correlated environmental noise, charge noise in silicon qubits leading to cross-correlations, or leakage and cosmic-ray events in superconducting qubits. If errors are positively correlated, the actual system error probability could be higher than $1-(1-p)^q$, making the derived upper bound on $q$ optimistic. This work provides a foundational bound under simplified assumptions; future research should aim to extend this framework by incorporating more sophisticated, correlated noise models. This could involve, for example, adapting cluster expansion approaches to decompose noise generators or developing alternative bounding techniques that account for specific correlation structures as discussed in \cite{F2025accuratehonest}, thereby enhancing the practical applicability of these bounds to specific hardware platforms.
In Section \ref{sec:exp_results}, by fixing the percentage $\beta$, we determine the minimum number of principal components, $D$, hence, the minimum number of qubits, $q$, we would require from the qPGA algorithm to efficiently encode the high-dimensional dataset in a $q$-qubit system capturing a percentage $\beta$ of variance as explained in Lemma \ref{lemma1}. We then experimentally demonstrate in Section \ref{sec:actual_device_results} that when we increase noise in the quantum system, we violate the upper bound set in Lemma \ref{lemma2} and obtain significant performance degradation during classification tasks. 

\section{Metric-based Assessment of qPGA in Data Pre-processing Stage for QML\label{sec:exp_results}}
In this section, we describe how we evaluate our proposed qPGA algorithm for encoding high-dimensional datasets into small-qubit systems as a feature extraction or dimensionality-reduction mechanism as described in Section \ref{sec:kpga_for_quantum}. We compare its performance against 2 quantum-based encoders used for dimensionality reduction. We also describe in this section the metrics we use to evaluate the performance of qPGA during the data pre-processing stages compared to the existing quantum-dependent methods. Then, we describe the metrics-based experiments performed, and present and discuss our results that involve qPGA as our proposed method in the data pre-processing stage for QML.
\subsection{Quantum-based Encoders as Competing Methods\label{subsec:competing_methods}} 

To benchmark our proposed qPGA algorithm for encoding high-dimensional classical datasets onto qubit systems by efficiently reducing the dimension of data, we compare its performance against two established methods. These methods are \textit{Quantum Autoencoder (QAE)} \cite{romero_quantum_2017, ma_compression_2023} and \textit{Hybrid Quantum Autoencoder (HAE)} \cite{tabi_ieee, srikumar_clustering_2022, sakhnenko_hybrid_2022}, and have been implemented in existing literature. These methods have demonstrated efficient compression of data using quantum-based autoencoders, and by comparing against these quantum-dependent methods, we aim to: (1) highlight the methodological innovation of qPGA, (2) demonstrate the resource efficiency and practicality of qPGA, which is classically computable--hence, not affected by the limitations of quantum hardware, and (3) demonstrate that no performance trade-off, due to constraints in current generation quantum computers, in the data pre-processing stage for QML is expected for qPGA as it is a quantum-independent method. 
In this work, we assume noise-free training and implementation of these quantum-dependent autoencoders to enable a fair comparison with our proposed qPGA algorithm.

In this work, we particularly adapt the QAE design and training procedure proposed by \cite{romero_quantum_2017}. Once trained, we use the Quantum Encoder (QE) to reduce the dimensionality of our original classical data, $\mathbf{X}\in \mathbb{R}^N$, which is then fed directly as quantum states of $\lceil\log_2D\rceil$-qubit systems involved in QML tasks, where $D$ represents the dimension of the latent space. We adopt the HAE structure proposed in \cite{sakhnenko_hybrid_2022}, where the Hybrid Quantum Encoder (HQE) consists of classical layers, with $N$ input and $D$ output nodes, followed by a $D$-qubit quantum circuit in its bottleneck layer to improve the latent space embeddings. Here, the embeddings are obtained after performing expectation measurements on the quantum circuit and hence lie in a $D$-dimensional Euclidean space. We provide further details of these 2 methods and their adaptations for this work in Appendix \ref{appendix0:methods}. A summary of these methods is presented in Table \ref{tab:method_details}.

\begin{table}[hbt!]
    \centering
    \caption{Summary of data pre-processing methods. qPGA uses normalized kernel feature mappings to project classical data onto a unit Hilbert sphere. Competing methods, QE and HQE, are also described. For all three methods, the input dimension is $N$, and output dimension is $D$.}
    \label{tab:method_details}
    \begin{tabular}{|>{\centering\arraybackslash}m{3cm}|p{10.5cm}|} 
        \hline
        \textbf{Methods} & \textbf{Detail} \\ \hline
        qPGA & \textit{Normalized Kernel Feature Maps}: Linear, Polynomial, RBF, Sigmoid \\ \hline
        Quantum Encoder (QE) & \textit{Model}: $\lceil\log_2N\rceil$-qubit quantum circuit with Amplitude Encoding \\ \hline
        Hybrid Quantum Encoder (HQE) & \textit{Model}: Classical linear layers followed by Phase Encoding onto $D$-qubit quantum circuit \\ \hline
    \end{tabular}
\end{table}

While techniques such as qPCA, QLDA, and QSFA, as described in Section \ref{lit_rev1}, have been proposed for quantum dimensionality reduction, we do not include them in our benchmarking for the following reasons. 
Unlike encoder-based methods such as QE and HQE, which explicitly define how classical data is embedded into quantum circuits (e.g., via amplitude or phase encoding), the above-mentioned algorithms assume oracle access to quantum states or rely on quantum RAM, which are not feasible on currently available hardware.
In contrast, qPGA directly addresses the encoding bottleneck by producing low-dimensional, quantum-compatible representations of high-dimensional classical data entirely within the classical domain. Additionally, to ensure a fair and implementable comparison, we focus on encoder-based techniques (QE and HQE) that have been demonstrated in QML pipelines and share the same objective as qPGA: producing low-dimensional representations of classical data that can be embedded into quantum circuits using standard encoding schemes with minimal qubit and gate overhead. Moreover, given the structure of QE and HQE (see Appendix \ref{appendix0:methods}), the outputs of the two competing quantum autoencoder-based methods are real valued, making them suitable for benchmarking qPGA. 
\subsection{Metrics to Assess qPGA Against Competing Methods \label{subsec:metrics}}
In this section, we describe four commonly used metrics for evaluating the performance of our proposed qPGA algorithm against QE and HQE. These metrics are \textit{Explained Variance} \cite{WOLD198737}, \textit{Co-ranking matrix} \cite{lee_quality_2009, lueks2011evaluatedimensionalityreduction}, \textit{Trustworthiness} and \textit{Continuity} \cite{stasis_semantically_2016}. We use the Explained Variance metrics to evaluate the performance of our qPGA algorithm for extracting features capturing a significant portion of the original dataset's variance in its latent space representations. The remaining metrics determine how well the local intrinsic structure of the high-dimensional original dataset is maintained in its low-dimensional representations. These metrics ensure that the latent space vectors effectively capture the characteristics of the high-dimensional dataset, thereby enabling its effective and accurate embedding onto small-qubit systems. 

Since standard Trustworthiness and Continuity metrics compare distance-based rank relationships in Euclidean space, we modified their computation for data lying on the UHS. Specifically, this applies to the low-dimensional representations obtained from qPGA and QE methods. To adapt these metrics, we modified the Trustworthiness and Continuity functions from the \code{scikit-learn} Python library to use geodesic distances instead of Euclidean distances when computing pairwise rankings on hyperspherical manifolds. The geodesic distances are computed using Eq. (\ref{eq:geodesic_distance}).

We describe these metrics in detail in Appendix \ref{appendix:metrics} and provide a summary in Table \ref{tab:metrics_table}. (*) indicate the modified versions of the Trustworthiness and Continuity metrics to account for geodesic distance instead of Euclidean distance, for low-dimensional embeddings of data obtained from the qPGA and QE methods for feature extraction. 
\begin{table}[hbt!]
    \centering
\caption{Metrics to assess the quality of feature extraction methods}
\label{tab:metrics_table}
    \begin{tabular}{|c|l|}
        \hline
         \textbf{Methods}&\textbf{Metrics}\\ \hline
         qPGA&Explained Variance, Co-ranking, Trustworthiness*, Continuity*\\ \hline
         QE&Trustworthiness*, Continuity*\\ \hline
        HQE&Trustworthiness, Continuity\\ \hline
    \end{tabular}
\end{table}

\subsection{Metric-based Experiments, Results, and Discussions \label{subsec:exp_setup}}
In this section, we describe the experiments conducted to assess qPGA against the competing methods during the data pre-processing stages for QML using the metrics presented in Section \ref{subsec:metrics}. 
First, we describe the datasets and how we pre-process them for our experiments assessing qPGA against the 2 competing methods. We then provide details of the experiments conducted and present the results and discussions. 
\subsubsection{Datasets and Data Pre-processing \label{subsec:datasets}}
The datasets we consider in this work are three publicly available benchmark image datasets, namely \textit{MNIST} \cite{mnist}, \textit{Fashion-MNIST (or FMNIST)} \cite{xiao2017fashionmnistnovelimagedataset}, and \textit{CIFAR-10} \cite{krizhevsky2010cifar}. We use these datasets to conduct experiments assessing our proposed qPGA algorithm against QE and HQE in the data pre-processing stages, and in downstream QML classification tasks (Section \ref{sec:classification}).
For the MNIST and FMNIST datasets, we extract 1,200 images and resize them to $8\times8$ grayscale images. For the CIFAR-10 dataset, we also extract 1200 images and convert these 3-channel images to 1-channel grayscale images of size 32 by 32. We do not reduce the size of CIFAR-10 images to prevent loss of information, given that it is a more complex dataset than MNIST and FMNIST.    
In our experiments, we treat these datasets as datasets of 1-dimensional vectors flattened from their 2-dimensional image representation. Thus, MNIST and FMNIST datasets are represented using 64-dimensional vectors, while CIFAR-10 consists of 1024-dimensional vectors. 
Also, we extract only two distinct classes from each of these datasets to make them suitable for end-to-end binary classification experiments in Section \ref{sec:classification}.
We provide further details of these datasets in Appendix \ref{appendix:datasets}.

\subsubsection{Experiments}
We first partitioned each dataset into 5 folds, allocating 80\% for training and 20\% for testing, and organized them into separate folders to ensure consistency of the datasets across all experiments. Next, we describe the experimental setup for our proposed qPGA algorithm, and QE and HQE as the competing methods.

\paragraph{\textit{Our Proposed Method--qPGA Algorithm}}
To map our input data to a UHS, we considered 4 normalized feature maps, namely: `Linear', `Polynomial (Poly) ($degree=3$) ', `Radial-basis function (RBF)', and `Sigmoid'. For the `Linear' feature map, we set the feature map as $\Phi(\textbf{X})=\frac{\textbf{X}}{||\textbf{X}||}$, which represents the normalized transformation of $\textbf{X}$, mapping the original data, $\textbf{X}$, onto a UHS in the input space.
Because it is non-trivial to compute exactly the remaining kernel feature maps, we used the \code{Nystroem} module, from the \code{kernel\_approximation} toolbox from the \code{scikit-learn} Python library for approximating the Polynomial ($degree=3$), RBF ($gamma=0.001$), and Sigmoid ($gamma=0.01$, $coef0=0$) kernel feature maps given input data $\textbf{X}$. To map the features to a UHS in RKHS, we normalized the resulting features to unit norm. 
Thus, these kernel feature maps project the $N$-dimensional input data to $\mathcal{S}^{N-1}$, which represents the surface of the UHS in an $N$-dimensional Hilbert space. They were then projected to $D$-dimensional UHS (latent space) as per the qPGA algorithm. In doing so, we mapped from $N$-dimensional inputs to a $D$-dimensional latent space representing the amplitude vector space of a $\lceil\log_2D\rceil$-qubit system. We used \code{geomstats}, an open-source Python package for statistical analyses and computations on nonlinear manifolds, to build our qPGA algorithm \cite{geomstats}.

For our experiments, we processed our input data from each fold using the qPGA algorithm and stored the latent representations separately. By setting $\beta=75\%$, we determined the number of principal components, $D$, required for capturing a cumulative explained variance $\gamma\geq\beta$ of the dataset variance. We did this by adding the explained variance captured by each ordered principal component to give the cumulative explained variance captured by the first $D$ principal components. $\beta=75\%$ serves as a good benchmark as it represents a significant portion of the dataset variance being represented by the principal components. Moreover, it characterizes a good representation of the high-dimensional dataset onto a UHS in the latent space. 
We then computed the co-ranking matrix, and trustworthiness and continuity for up to $k=50$ neighbors, on 960 dataset samples, obtained from the training samples corresponding to the second fold. We present these results in the next section. It is important to note that the choice of kernel and its parameters impacts qPGA's performance, acting as crucial hyperparameters. The specific parameters for the kernels used in this study (e.g., $degree$ for Polynomial, $gamma$ for RBF, and Sigmoid) were chosen based on common default values and preliminary exploratory analysis. This work empirically explores four common kernels (Linear, Polynomial, RBF, and Sigmoid), providing valuable insights. Future investigations could focus on developing heuristics or automated methods for optimal kernel selection based on dataset properties (e.g., linearity, cluster shapes, intrinsic dimensionality) to further enhance qPGA's ease of use and maximize its effectiveness. 

\paragraph{\textit{Competing Method 1--Quantum Encoder (QE)}} 
We normalized our input data of dimension $N$ and loaded them onto the quantum autoencoder (QAE) using the amplitude embedding block as shown in the Appendix Fig. \ref{fig:quantum_encoder}. For the MNIST and FMNIST datasets, since $N=64$ and $D=4$, the number of input qubits is $n_{trash}+n_{latent}=\lceil\log_2N\rceil=6$, $n_{latent}=\lceil\log_2D\rceil=2$ and $n_{trash} = n_{ref} = 4$, and including 1 auxiliary qubit, the QAE consists of 11 qubits, which is exactly the one shown in Fig. \ref{fig:quantum_encoder}, having 36 parameters to optimize. For the CIFAR-10 dataset, since we set $N=1024$ and $D=16$, the QAE consists of $10+6+1=17$ qubits. The number of latent qubits is $n_{latent}=4$. The ansatz used in this case is still the \code{RealAmplitudes} ansatz, but scaled for 10 qubits and hence has 60 trainable parameters. We built the QAE using the Pennylane Python package \cite{bergholm2022pennylaneautomaticdifferentiationhybrid} on the \code{default-qubit} simulator with the \code{backprop} backpropagation method, making it computationally efficient to simulate \cite{JoshIzaac2020}. We trained the QAE for 50 epochs using the \code{Adam} optimizer with a learning rate of $10^{-3}$ against the fidelity loss function as described in Appendix \ref{subsec:method2}. 

After training on the first fold, we extracted and applied the trained Quantum Encoder (QE) to compress the input training data for the next fold without further retraining. Given a quantum state expressed as \( |\psi\rangle = \sum_i \alpha_i |i\rangle \), we applied the built-in PennyLane function \texttt{probs} to compute the measurement probabilities \( |\alpha_i|^2 \) over the computational basis states. To recover the real-valued magnitudes of the amplitudes, we then took the square root of these probabilities, obtaining \( |\alpha_i| \) for each basis state. While this procedure discards the complex phase information of the quantum state, it yields real-valued, unit-normed latent vectors consistent with the outputs of the qPGA algorithm, which also generates real-valued amplitude vectors. This ensures an appropriate alignment of the representation spaces, enabling a fair comparison based on our evaluation metrics. We then computed the trustworthiness and continuity scores between the second fold’s input training data and its latent representation. The procedure was repeated across all datasets, and the results are presented in the next section.

\noindent
\paragraph{\textit{Competing Method 2--Hybrid Quantum Encoder (HQE)}}
As described in Appendix \ref{subsec:method3}, the HQE consists of 2 classical neural network layers with $N$ input nodes and $D$ output nodes, followed by a $D$-qubit quantum circuit whose $D$-dimensional output represents the latent representation of the $N$-dimensional input. The circuit is as shown in the Appendix Fig. \ref{fig:circuit_hqe}. The $D$-dimensional expectation measurements of the quantum circuit are then fed to the decoder model, which consists of 2 classical layers with $D$-dimensional input nodes and $N$-dimensional output nodes. At the end of each classical layer, we use the \code{Tanh} activation function. 
For the MNIST and FMNIST datasets, we set $N=64$, corresponding to a 64-dimensional vector representing the input images, and $D=4$, corresponding to a 4-dimensional latent space. For the CIFAR-10 dataset, $N=1024$ and $D=16$.

We trained the HAE for 20 epochs using the \code{Adam} optimizer with a learning rate of $10^{-3}$ and the \code{MSELoss} as loss function. We trained the HAE on the first fold of each dataset. Then, we extracted and ran the trained encoder (HQE) to obtain a compressed representation of the original data from the second fold. We computed the trustworthiness and continuity as in the case of qPGA and QE, and the results are presented next. 

\subsubsection{Results and Discussions\label{subsec:results_part1}} 
In this section, we present and discuss our experimental results corresponding to the metrics computation assessing the performance of our proposed qPGA algorithm against QE and HQE during the data pre-processing stages for QML. 

\begin{figure}[hbt!]
   \begin{subfigure}{0.49\textwidth}
    \centering
        \includegraphics[width=\textwidth]{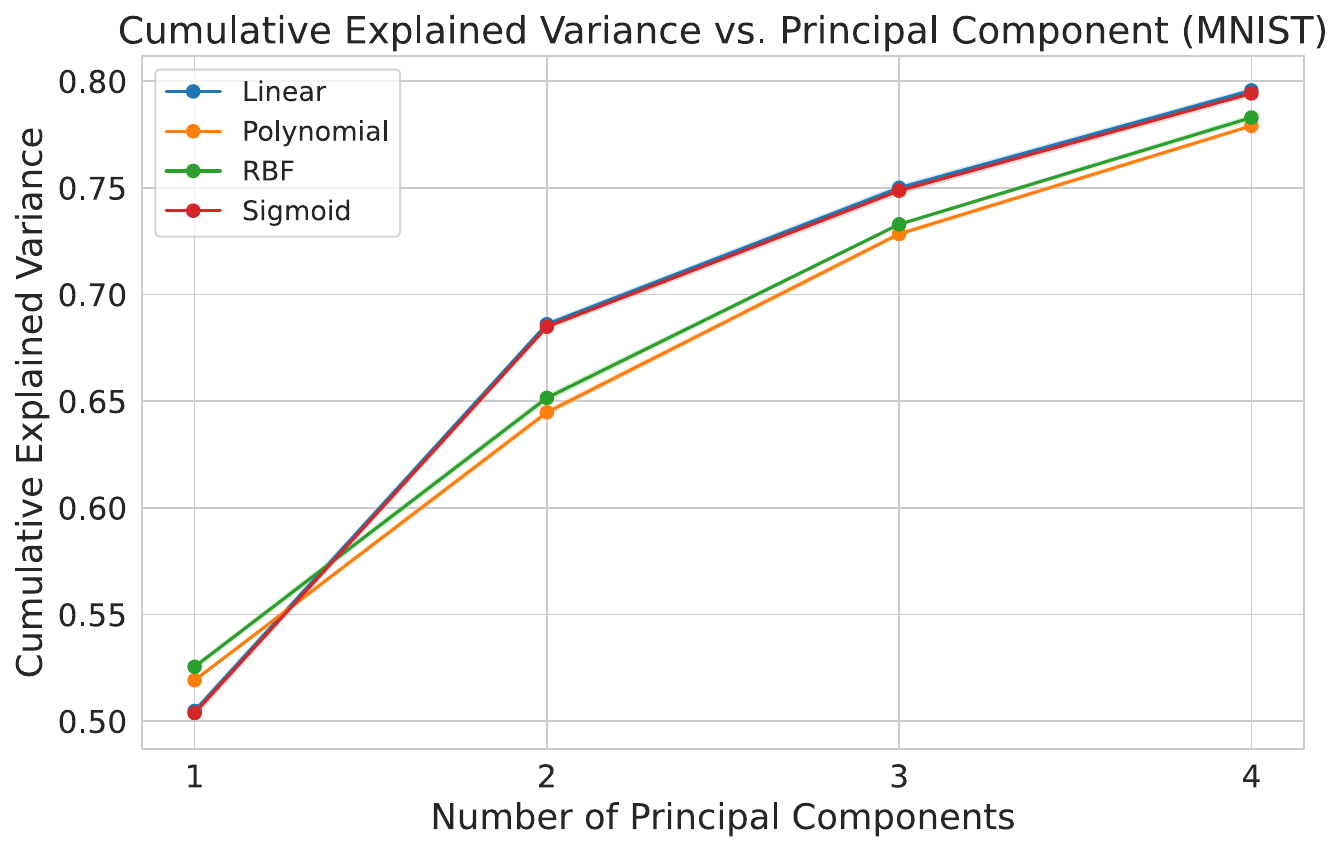}
        \caption{Cumulative Explained Variance on MNIST}
        \Description{Cum EV for MNIST vs PCs}
        \label{fig:mnist_cum_evr}
    \end{subfigure}
    \medskip
       \begin{subfigure}{0.49\textwidth}
    \centering
        \includegraphics[width=\textwidth]{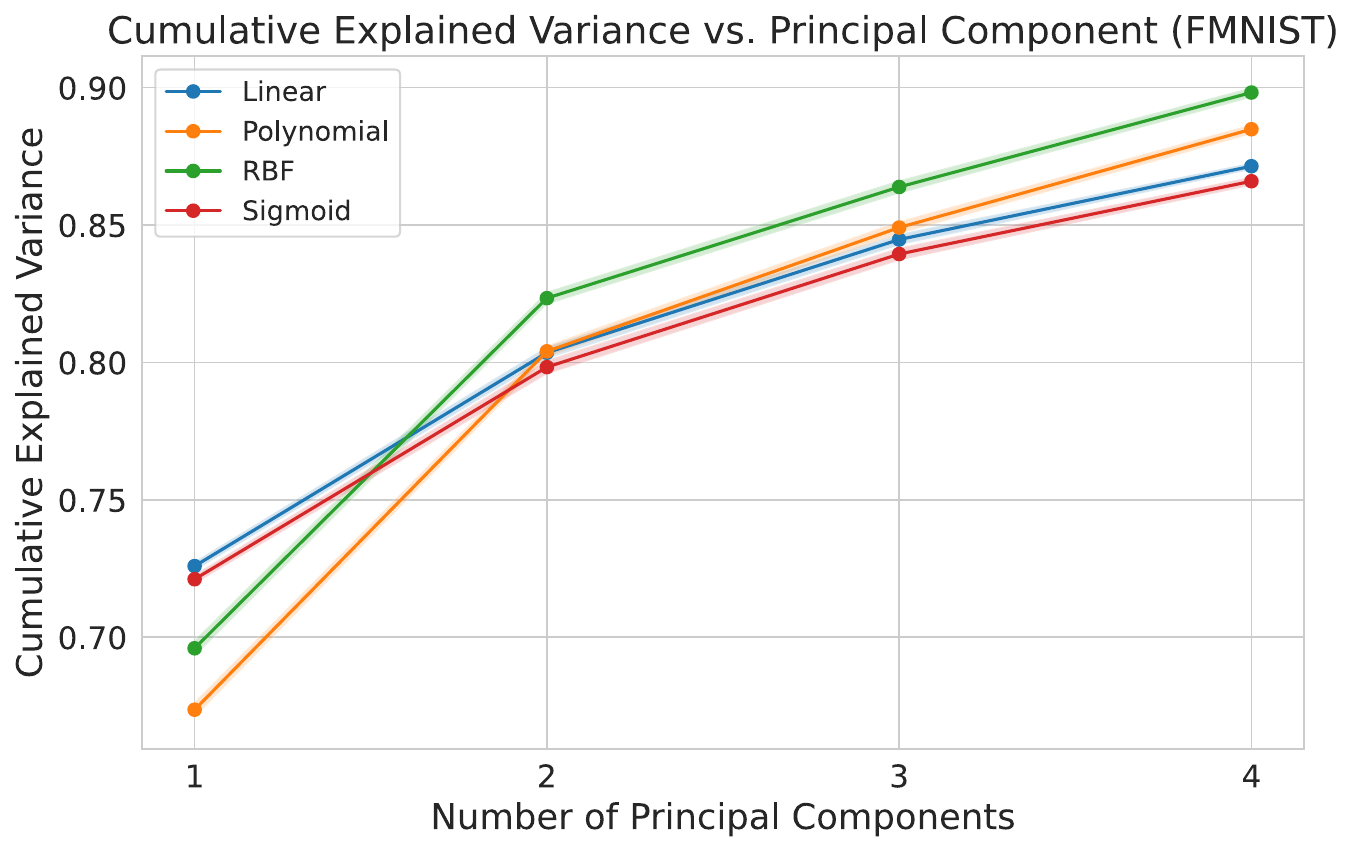}
        \caption{Cumulative Explained Variance on Fashion-MNIST}
        \Description{Cum EV for FMNIST vs PCs}
        \label{fig:fmnist_cum_evr}
    \end{subfigure}
    \medskip
    \begin{subfigure}{\textwidth}
    \centering
        \includegraphics[width=0.49\textwidth]{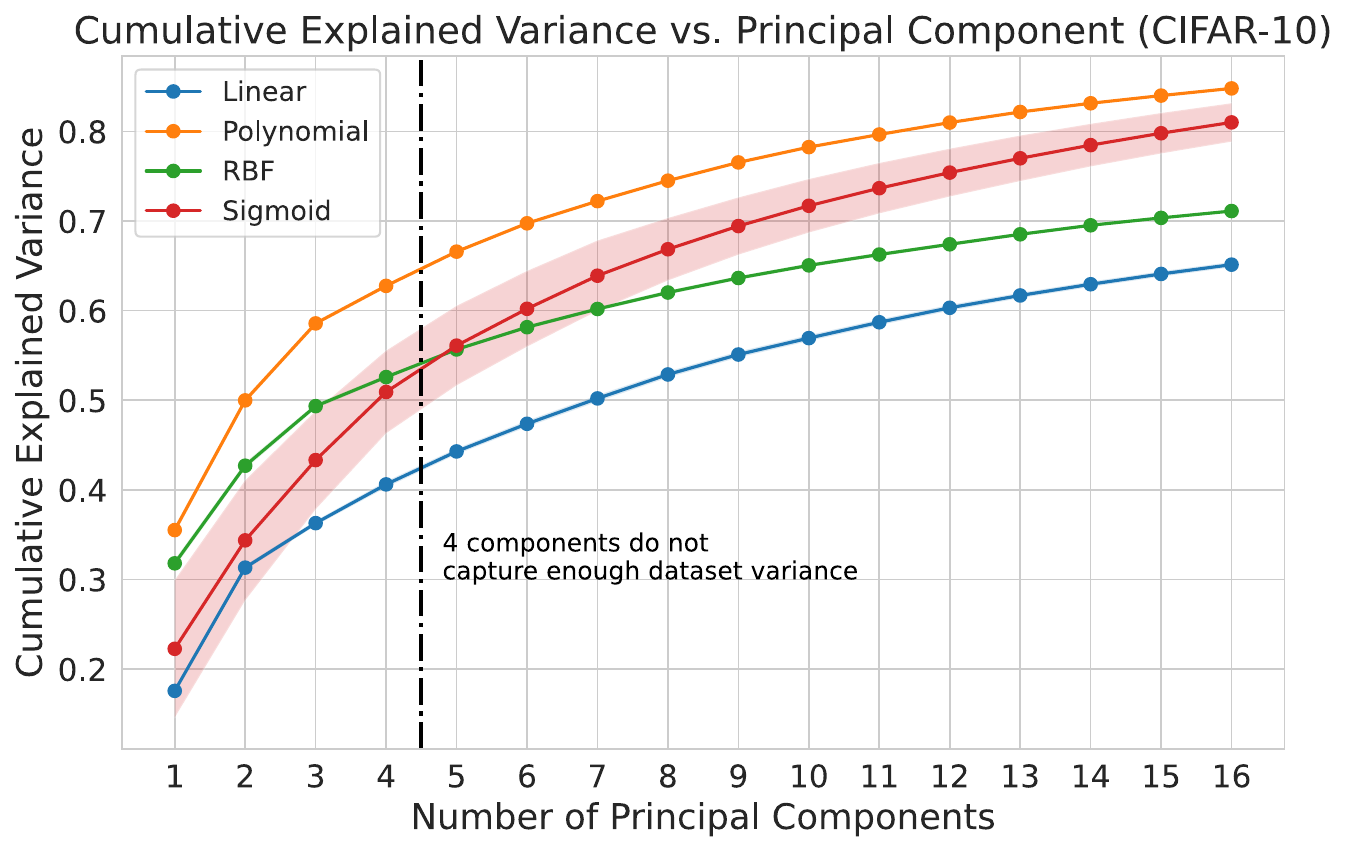}
        \caption{Cumulative Explained Variance on CIFAR-10}
        \Description{Cum EV for CIFAR-10 vs PCs}
        \label{fig:cifar10_cum_evr}
    \end{subfigure}
    \medskip
    \caption{5-fold Mean Cumulative explained variance using the first $D$ principal components from qPGA with different kernel feature maps. (a), (b) For MNIST and FMNIST, $D=4$ is sufficient to explain $\beta=75\%$ of variance. (c) For CIFAR-10, $D=16$ principal components are needed due to the dataset's higher complexity. The shaded regions indicate the variance across multiple runs.}
    \Description{Results for Cumulative Explained Variance (Cum EV) against principal components (PCs) for 3 datasets}
    \label{fig:explained_variance_comparisons}
\end{figure}
\noindent
\paragraph{\textit{Explained Variance}.}
For the qPGA algorithm, we computed the cumulative explained variance, $\gamma$, captured by the first $D$ principal components capturing a combined variance of at least $\beta=75\%$ of the original dataset variance. For the MNIST and FMNIST datasets, where the dimension of the original dataset is $N=64$, $D=4$ principal components were sufficient to achieve $\gamma\geq\beta$,  while for the CIFAR-10 dataset, where $N=1024$ and $D=16$ principal components were required to satisfy $\gamma\geq\beta=75\%$ for only the polynomial and sigmoid kernel feature maps. However, we did not manage to achieve the set $\beta=75\%$, from $D=16$ principal components for the other kernel feature maps. Increasing $D$ would most probably satisfy $\gamma\geq\beta=75\%$ for all kernel feature maps. However, that would also result in an increase in the number of qubits required to embed the latent space as explained in Section \ref{theory} Lemma \ref{lemma1}. 

We display these findings in Fig. \ref{fig:explained_variance_comparisons}, which shows the mean and variance of the cumulative explained variance obtained when using the qPGA algorithm during the data pre-processing stages on the training set from each of the 5 folds of the dataset. 
From these results, we can deduce that the CIFAR-10 dataset is a more complex dataset compared to MNIST and FMNIST; a larger number of principal components was required to capture a cumulative explained variance $\gamma\geq\beta=75\%$ of the dataset variance. This also implies that for the CIFAR-10 dataset, we require a larger qubit system to embed enough variance from the dataset in its latent representation.

In contrast, for MNIST and FMNIST, only $D=4$ principal components were enough to achieve $\gamma\geq\beta=75\%$. 
The high cumulative explained variance obtained by the first 4 principal components for the MNIST and FMNIST datasets also implies that the dataset has redundancy or strong correlations among features, allowing for fewer components to represent the data on a 4-dimensional UHS effectively. We also note that the explained variance captured by the principal components is dependent on the kernel feature map we apply to our inputs to our qPGA algorithm. Hence, the feature map to use acts as a hyperparameter that needs to be tuned in the qPGA algorithm. 

Referring to Lemma \ref{lemma1} in Section \ref{theory}, the number of qubits, $q$, required has a lower bound dependent on $\beta$, the proportion of variance we want to capture. By Lemma \ref{lemma2}, the number of qubits, $q$, has an upper bound that is dependent on the maximum permissible qubit error, $\mathcal{P}_{max}$. This implies that we cannot have a large number of qubits due to increasing qubit error with increasing $q$, even though a large $q$ would allow us to capture a larger portion of the dataset variance. We show in Section \ref{sec:actual_device_results} that the upper bound is violated when the noise level characterized by the probability of qubit error $P(error)=p$, increases. This implies that we should reduce the number of qubits to stay within bounds; however, a small number of qubits would encode fewer principal components, capturing less dataset variance, hence poorly representing the dataset. \textit{The representation power of the qPGA algorithm is thus constrained by the increased noise levels}.

\begin{figure}[!htpb]
   \begin{subfigure}{0.49\textwidth}
    \centering
        \includegraphics[width=\textwidth]{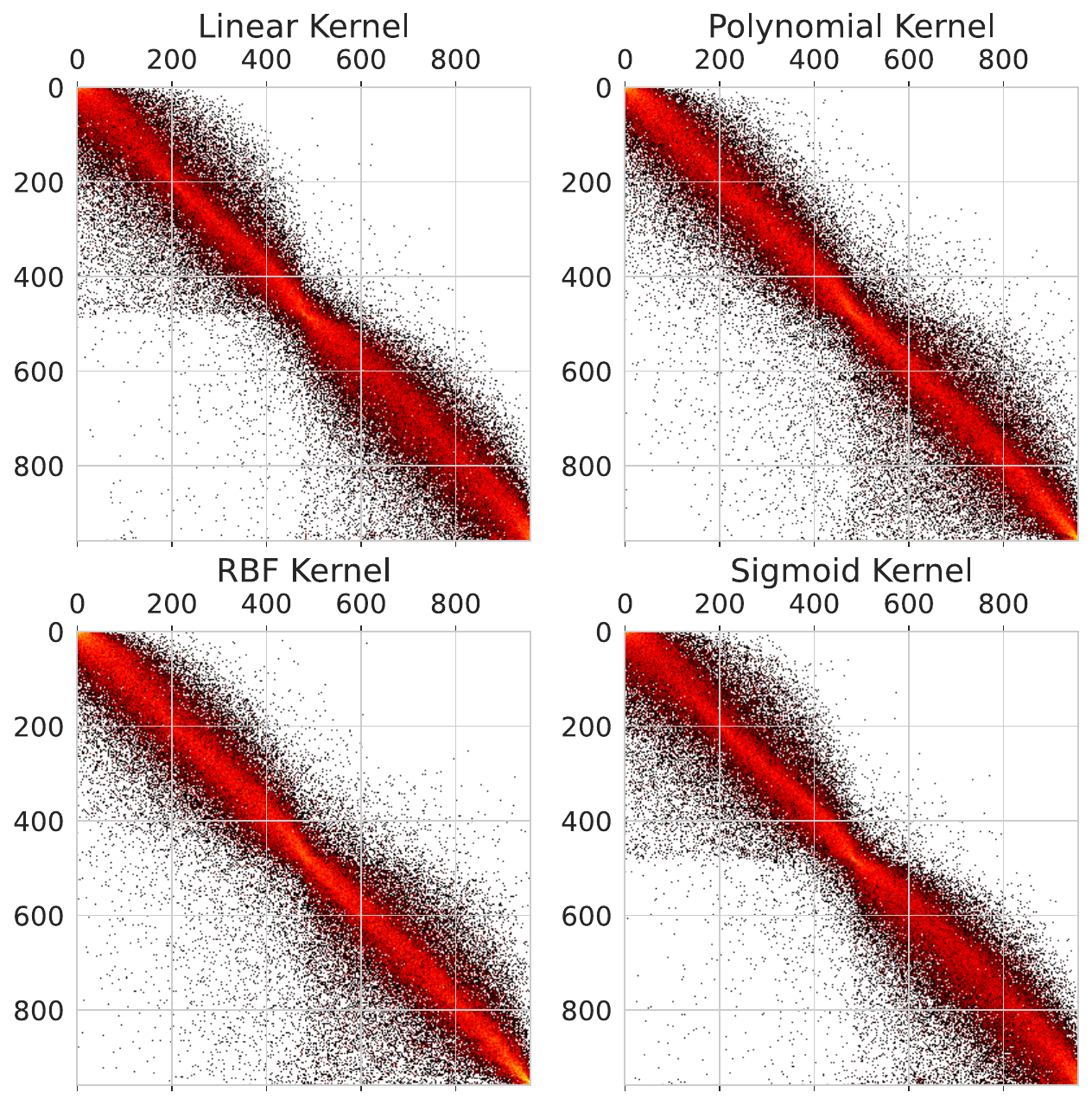}
        \caption{Co-ranking matrices - MNIST dataset ($D=4$)}
        \Description{MNIST CORANKING}
        \label{fig:mnist_corank}
    \end{subfigure}
    \medskip
    \begin{subfigure}{0.49\textwidth}
        \centering
        \includegraphics[width=\textwidth]{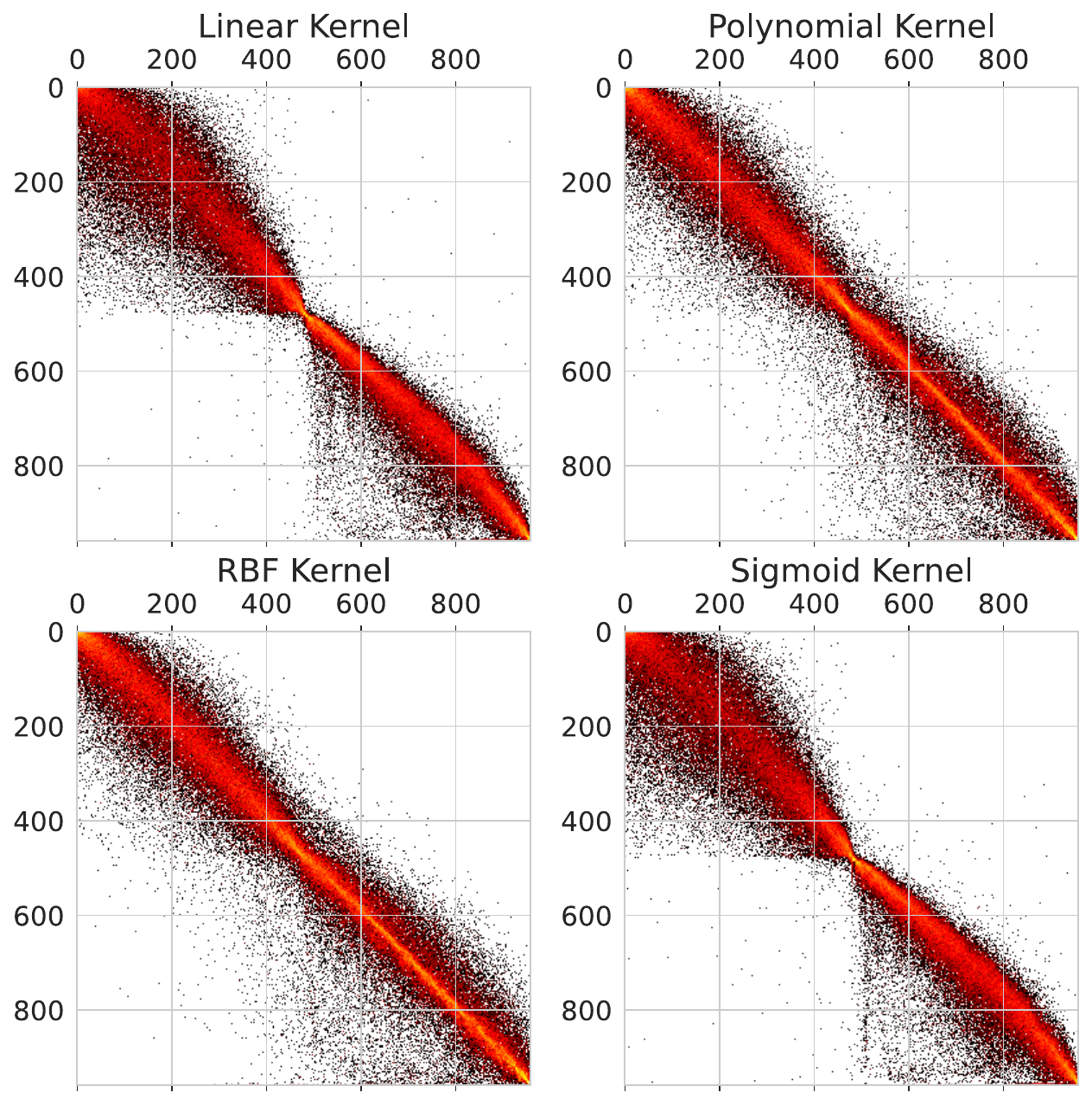}
        \caption{Co-ranking matrices - FMNIST dataset ($D=4$)}
        \Description{FMNIST CORANKING}
        \label{fig:fmnist_corank}
    \end{subfigure}
    \medskip
    \begin{subfigure}{\textwidth}
        \centering
        \includegraphics[width=0.49\textwidth]{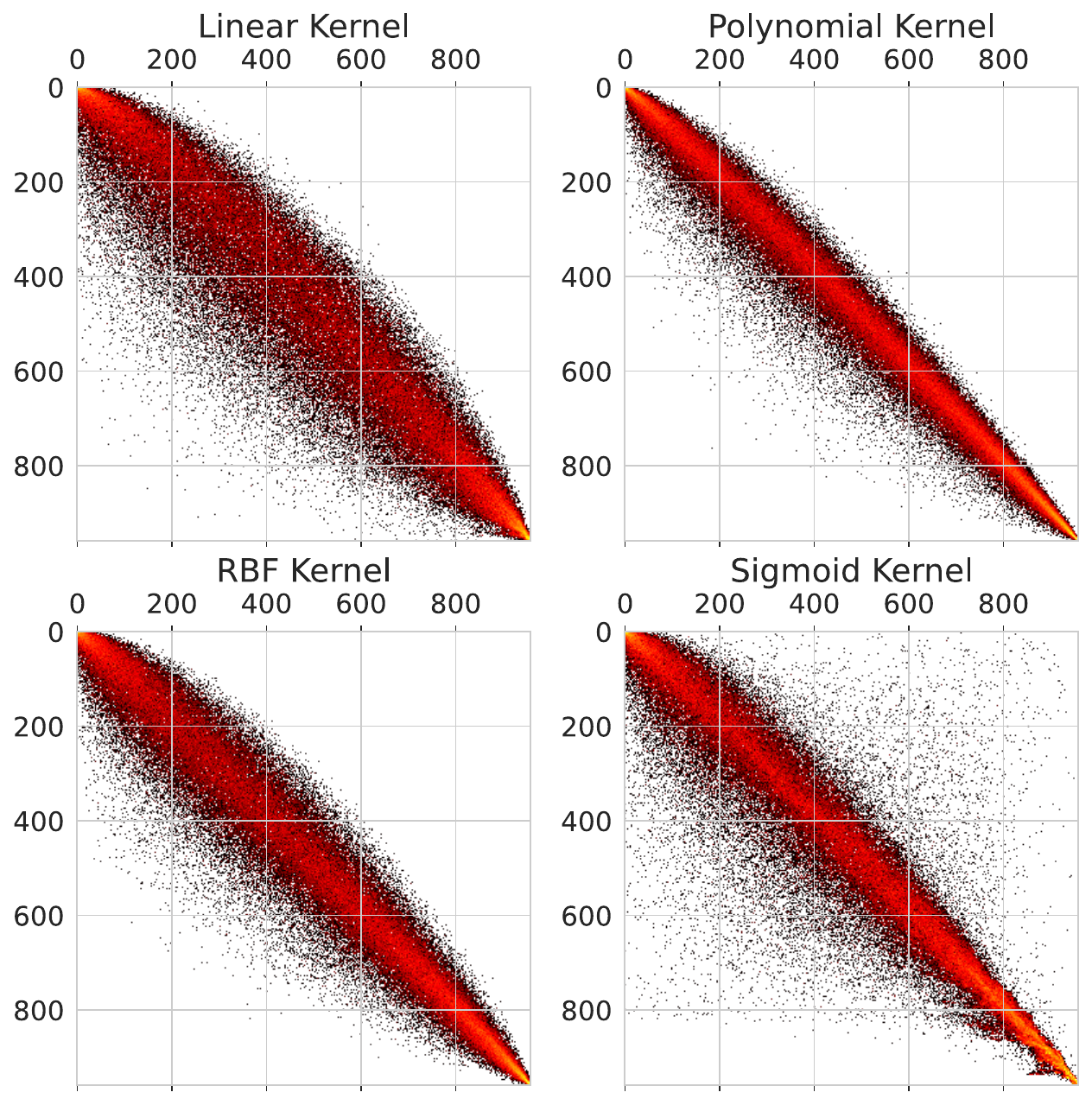}
        \caption{Co-ranking matrices - CIFAR-10 dataset ($D=16$)}
        \Description{CIFAR-10 CORANKING}
        \label{fig:cifar10_corank}
    \end{subfigure}
    \medskip
    \caption{Co-ranking matrices comparing high and low-dimensional data processed by qPGA with four different kernel feature maps. The sharper the red clusters along the diagonal, the better the rank relationships are preserved. For MNIST and FMNIST, $D=4$ components maintain key structures of the dataset, while CIFAR-10 requires $D=16$ for a sharp diagonal distribution.}
    \Description{CORANKING PLOTS for 3 datasets}
    \label{fig:coranking}
\end{figure}

\noindent
\paragraph{\textit{Co-ranking matrix}.}
The co-ranking matrices compare the rankings of pairwise distances between high-dimensional data points and their low-dimensional projections obtained by the qPGA algorithm with 4 different kernel feature maps. We present a visual representation of the co-ranking matrices in Fig. \ref{fig:coranking}. We use the \code{coranking} Python library to compute the co-ranking matrices \cite{samuel_k_git}. 

The co-ranking matrices indicate that, for all datasets, the majority of rank relationships between data points are preserved when reducing dimensionality using qPGA. This result demonstrates the effectiveness of qPGA in retaining the essential structural information during dimensionality reduction.
In co-ranking matrices, a sharper concentration of values along the diagonal reflects better preservation of rank relationships between data points. For MNIST and Fashion-MNIST datasets, only $D=4$ principal components were sufficient to produce a visibly sharp diagonal distribution, indicating improved alignment between the high-dimensional data and its latent representation. However, for the CIFAR-10 dataset, we required $D=16$ to achieve a sharp distribution along the diagonal of the co-ranking matrices. A visual representation of the co-ranking matrices for CIFAR-10 dataset when $D=\{4, 16, 32\}$ is presented in Appendix \ref{appendix1} illustrating the need for more principal components to improve rank relationships between the high-dimensional data and their latent space representation, characterized by a sharper distribution along the diagonal of the co-ranking matrices. We note here that a trade-off between the encoding circuit depth of our qubit system and the efficient encoding of our high-dimensional dataset is expected (as amplitude encoding requires an exponentially deep encoding circuit when the number of qubits increases). However, we do not tackle this trade-off in this work as we solely focus on the encoding efficiency aspect of qubit systems.

\begin{figure}[hbt!]
   \begin{subfigure}{\textwidth}
    \centering
        \includegraphics[width=0.49\textwidth]{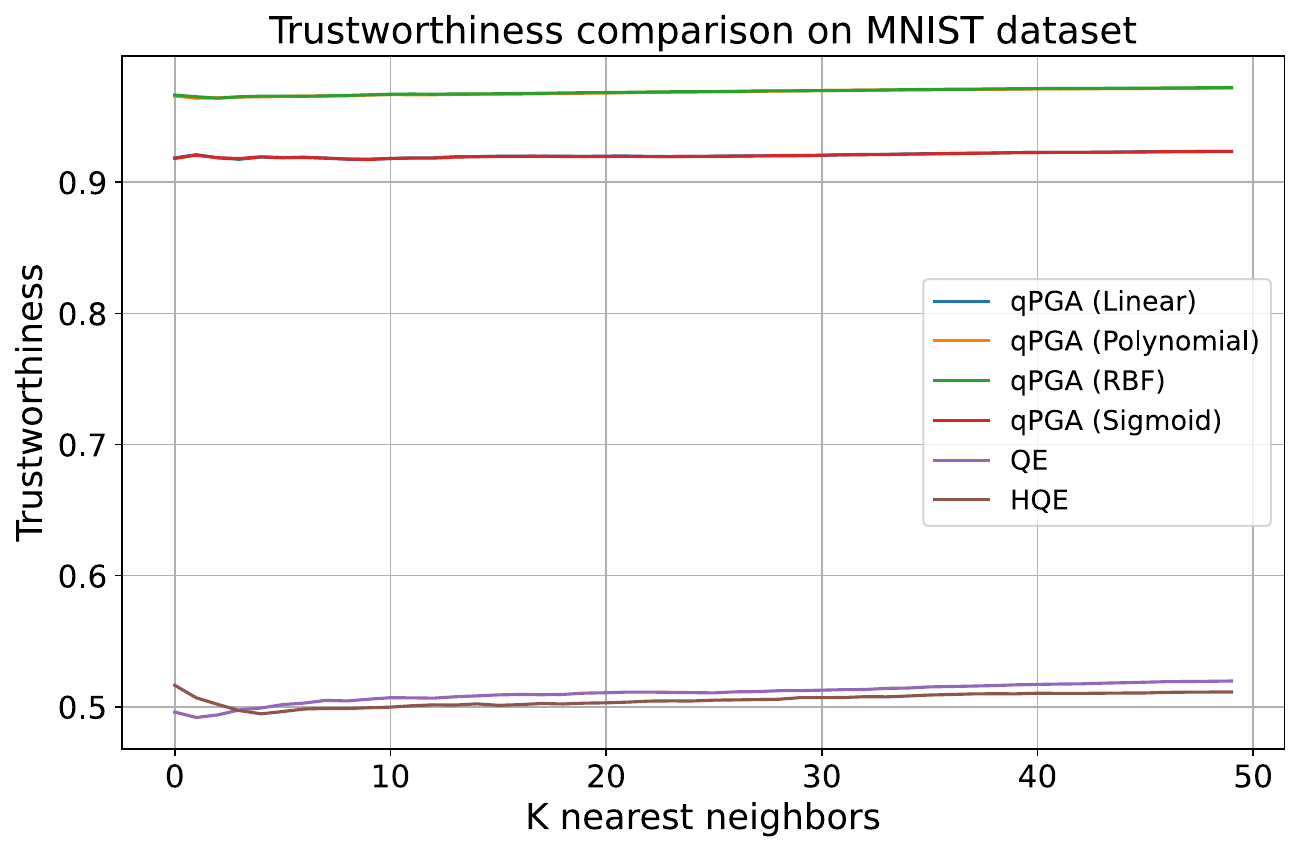}
        \includegraphics[width=0.49\textwidth]{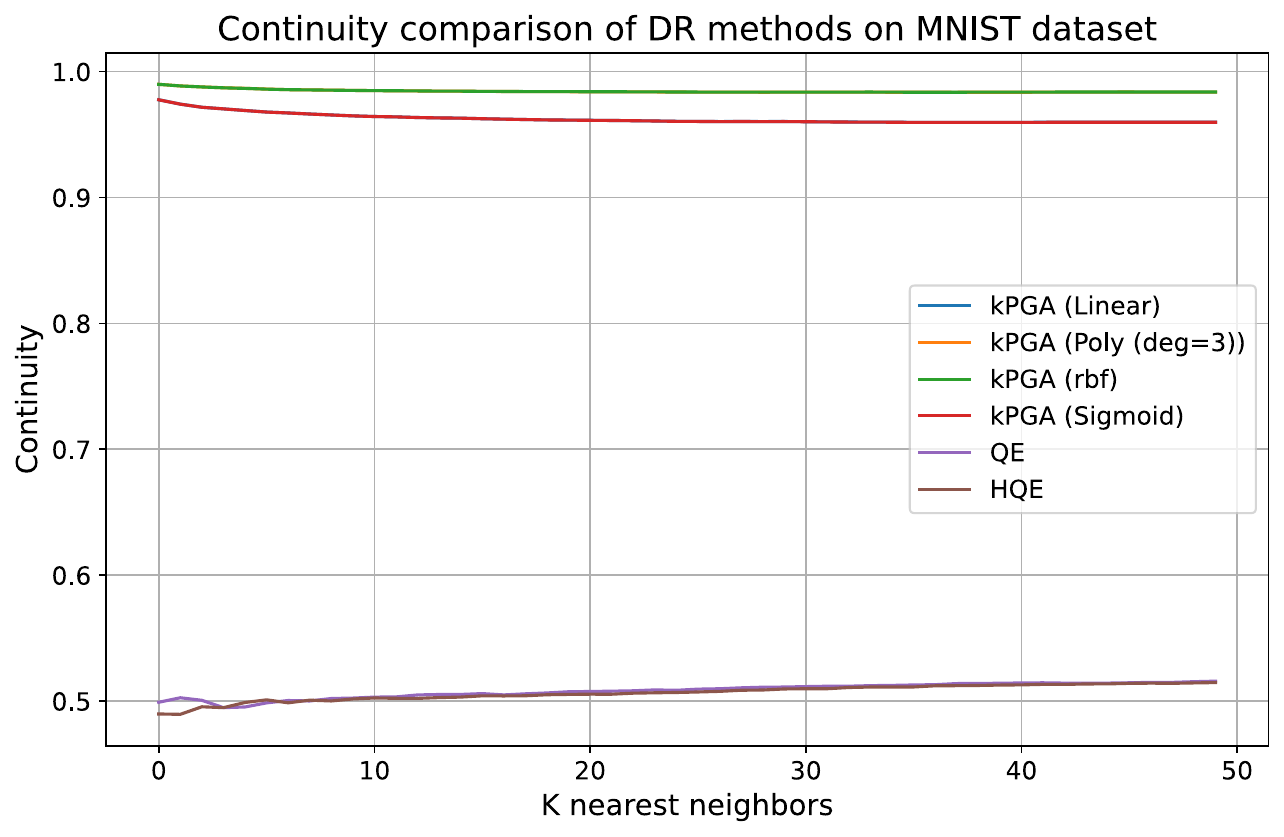}
        \caption{Trustworthiness and Continuity comparison of feature extraction methods on MNIST dataset}
        \Description{TNC MNIST}
        \label{fig:mnist_tnc}
    \end{subfigure}
    \medskip
    \begin{subfigure}{\textwidth}
    \centering
        \includegraphics[width=0.49\textwidth]{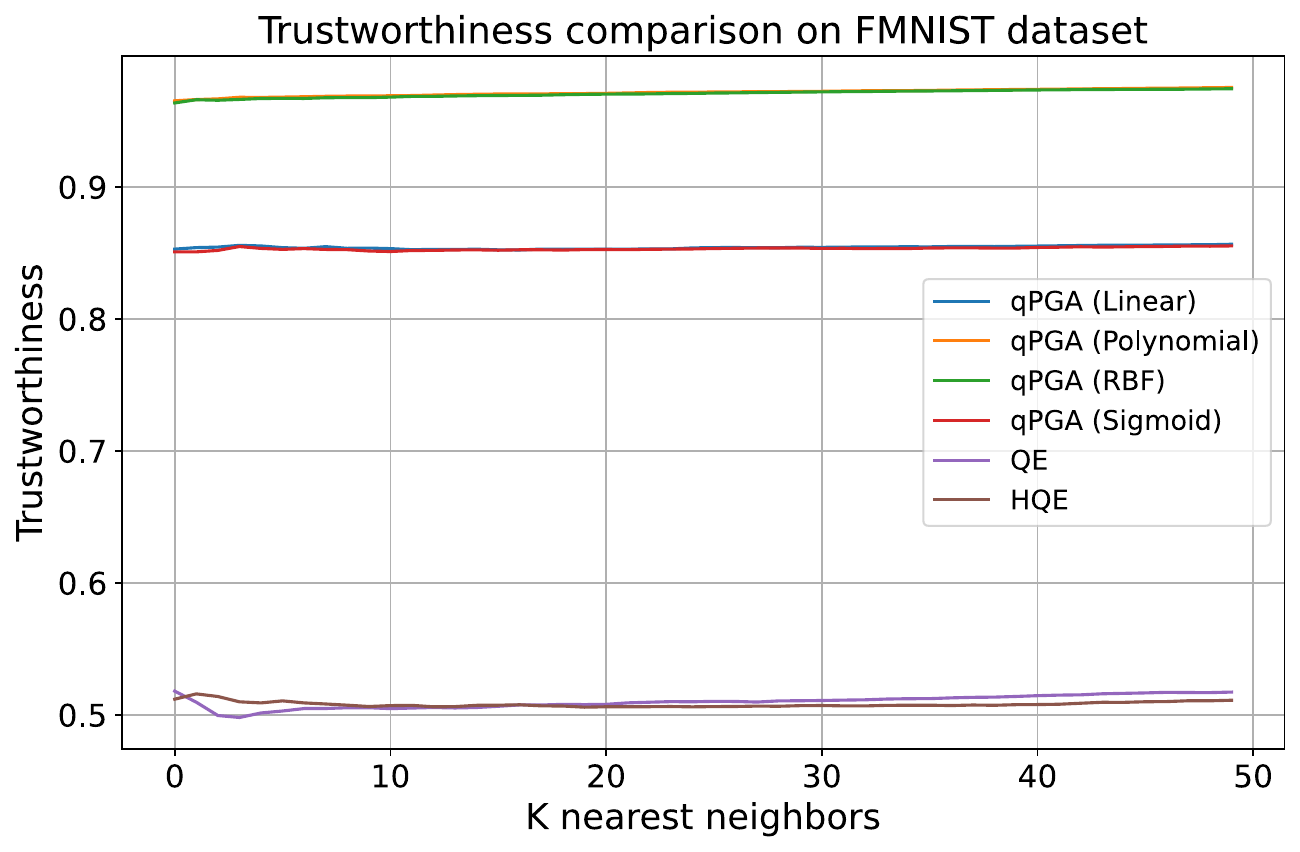}
        \includegraphics[width=0.49\textwidth]{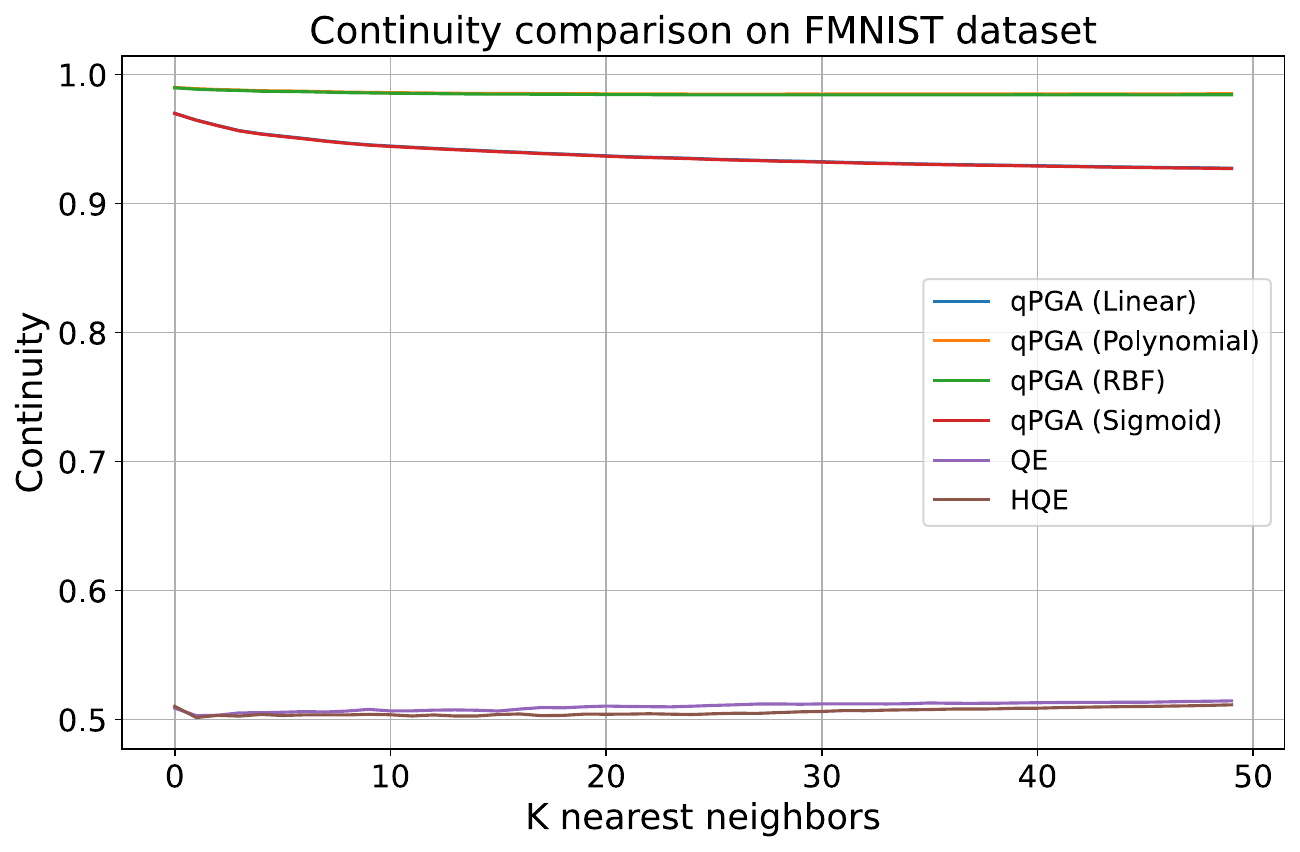}
        \caption{Trustworthiness and Continuity comparison of feature extraction methods on FMNIST dataset}
        \Description{TNC FMNIST}
        \label{fig:fmnist_tnc}
    \end{subfigure}
    \medskip
       \begin{subfigure}{\textwidth}
    \centering
        \includegraphics[width=0.49\textwidth]{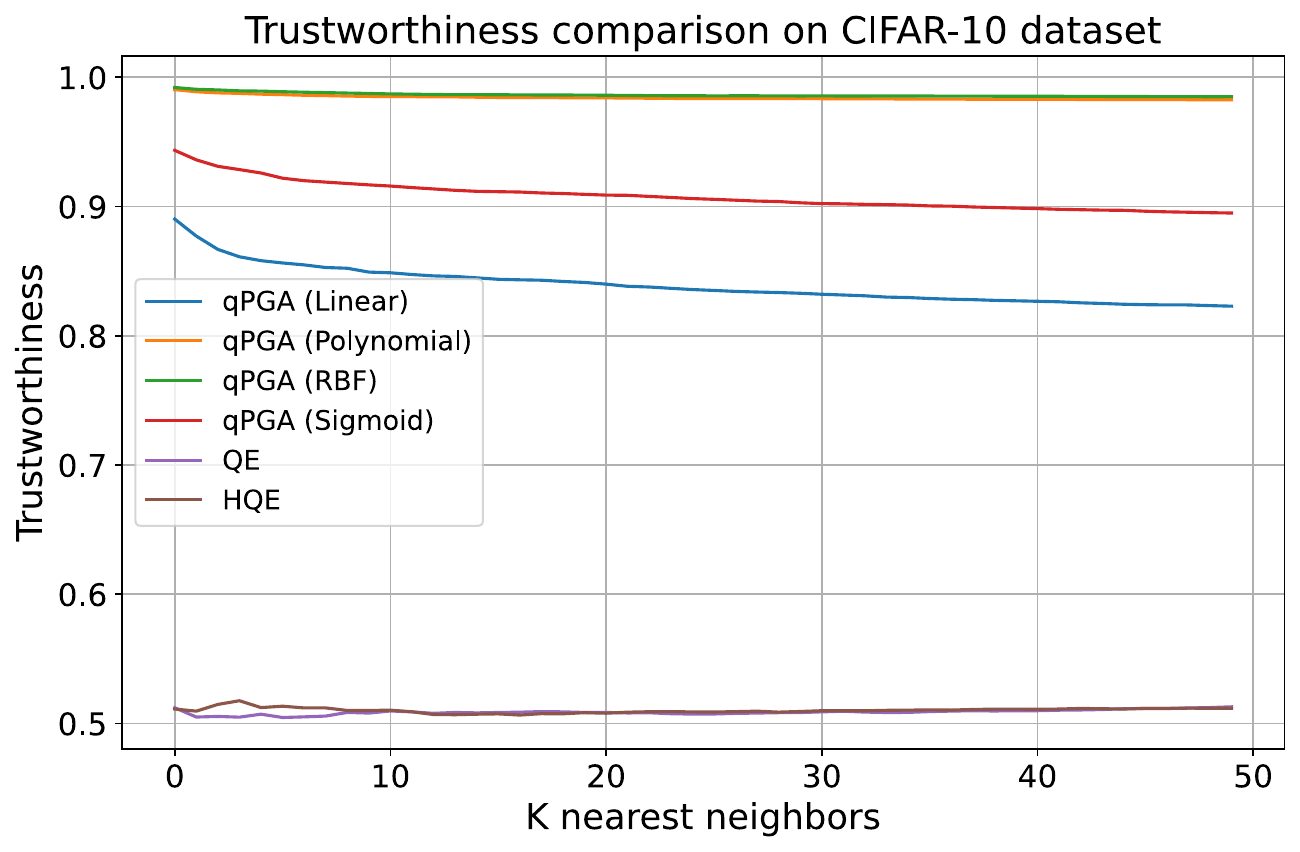}
        \includegraphics[width=0.49\textwidth]{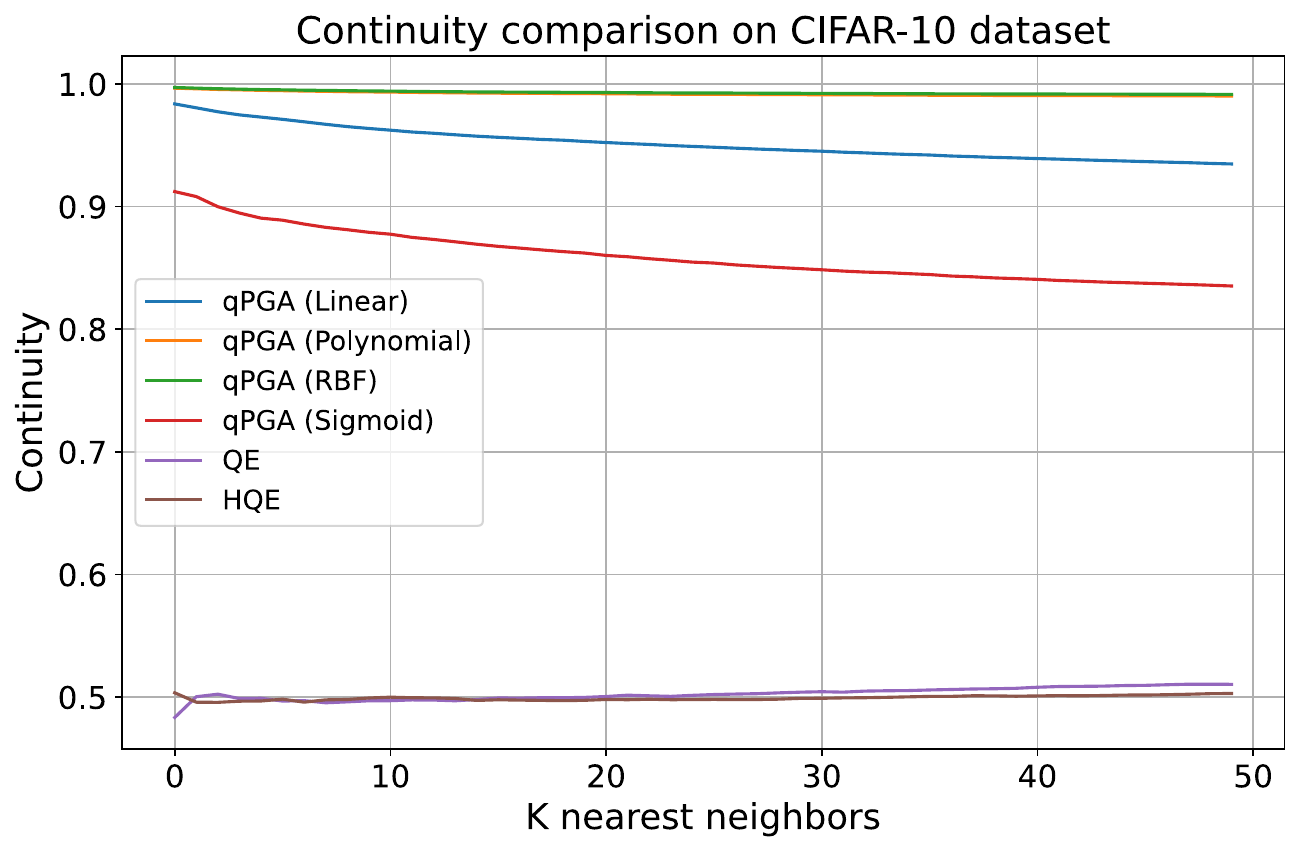}
        \caption{Trustworthiness and Continuity comparison of feature extraction methods on CIFAR-10 dataset}
        \Description{TNC CIFAR-10}
        \label{fig:cifar10_tnc}
    \end{subfigure}
    \medskip
    \caption{Trustworthiness and Continuity plots for qPGA (with 4 different kernel feature maps), Quantum Encoder (QE), and Hybrid Quantum Encoder (HQE) on MNIST, FMNIST, and CIFAR-10 datasets. qPGA maintains the local neighborhood structure better than the competing methods.}
    \Description{TNC plots for 3 datasets}
    \label{fig:tnc_comparisons}
\end{figure}

\noindent
\paragraph{\textit{Trustworthiness and Continuity.}}
Next, we compute the Trustworthiness and Continuity between the high-dimensional input data and their low-dimensional representations using the qPGA algorithm (with 4 different kernel feature maps), QE, and HQE as 3 distinct feature extraction methods. The results are shown in Fig. \ref{fig:tnc_comparisons}.

We can deduce that qPGA better preserves the local neighborhood structure of the data points compared to QE and HQE, as competing methods. In other words, points near each other in the high-dimensional space also appear near each other after dimensionality-reduction, as explained by the Trustworthiness plots -- and vice-versa for the Continuity plots. These observations are expected, as while autoencoders can approximately preserve local neighborhoods, especially if trained carefully, e.g., contractive autoencoders, qPGA is a method that directly preserves neighborhood structure by design.

\noindent
\textit{Overall, these metrics show that our proposed qPGA algorithm effectively reduces the dimensionality of our high-dimensional data using only a few principal components while retaining a major proportion of the dataset variance, as demonstrated by the cumulative explained variance plots in Fig. \ref{fig:explained_variance_comparisons}. Moreover, the qPGA algorithm for feature extraction maintains the intrinsic neighborhood structure of the original data in its latent representations better than the Quantum Encoder and Hybrid Quantum Encoder methods during the data pre-processing stages for QML tasks. This is shown by the co-ranking matrices (Fig. \ref{fig:coranking}) and the Trustworthiness and Continuity metrics (Fig. \ref{fig:tnc_comparisons}).} 

\noindent
We conducted an additional empirical study on the invertibility of our qPGA algorithm to retrieve the original datasets as opposed to QE and HQE as competing methods. In the next section, we describe our experiments and present our findings involving the invertibility of our qPGA algorithm, and hence its robustness to the reconstruction of original data from its latent space representations. This study is crucial from an adversarial attack perspective. 

\section{\label{sec:reconstruction}Invertibility of qPGA vs. Competing Methods}
In this section, we examine the invertibility of our proposed qPGA algorithm during the data pre-processing stages for QML tasks. Invertibility from latent space representations is a concern in adversarial reconstruction attack scenarios \cite{salem2020updates, hevish_HQSL}. If a feature extraction method is easily invertible, it is prone to data privacy leakage and reconstruction attacks \cite{hevish_HQSL} when an adversary has access to the latent-space representations. Here, we investigate how easily the features extracted from the qPGA algorithm can be inverted to the original data compared to the competing encoder-based methods.
We compute the mean square error (MSE) between the original and reconstructed data to assess the invertibility of our qPGA algorithm compared to the competing methods.
\begin{figure}[hbt!]
   \begin{subfigure}{0.49\textwidth}
    \centering
        \includegraphics[width=\textwidth]{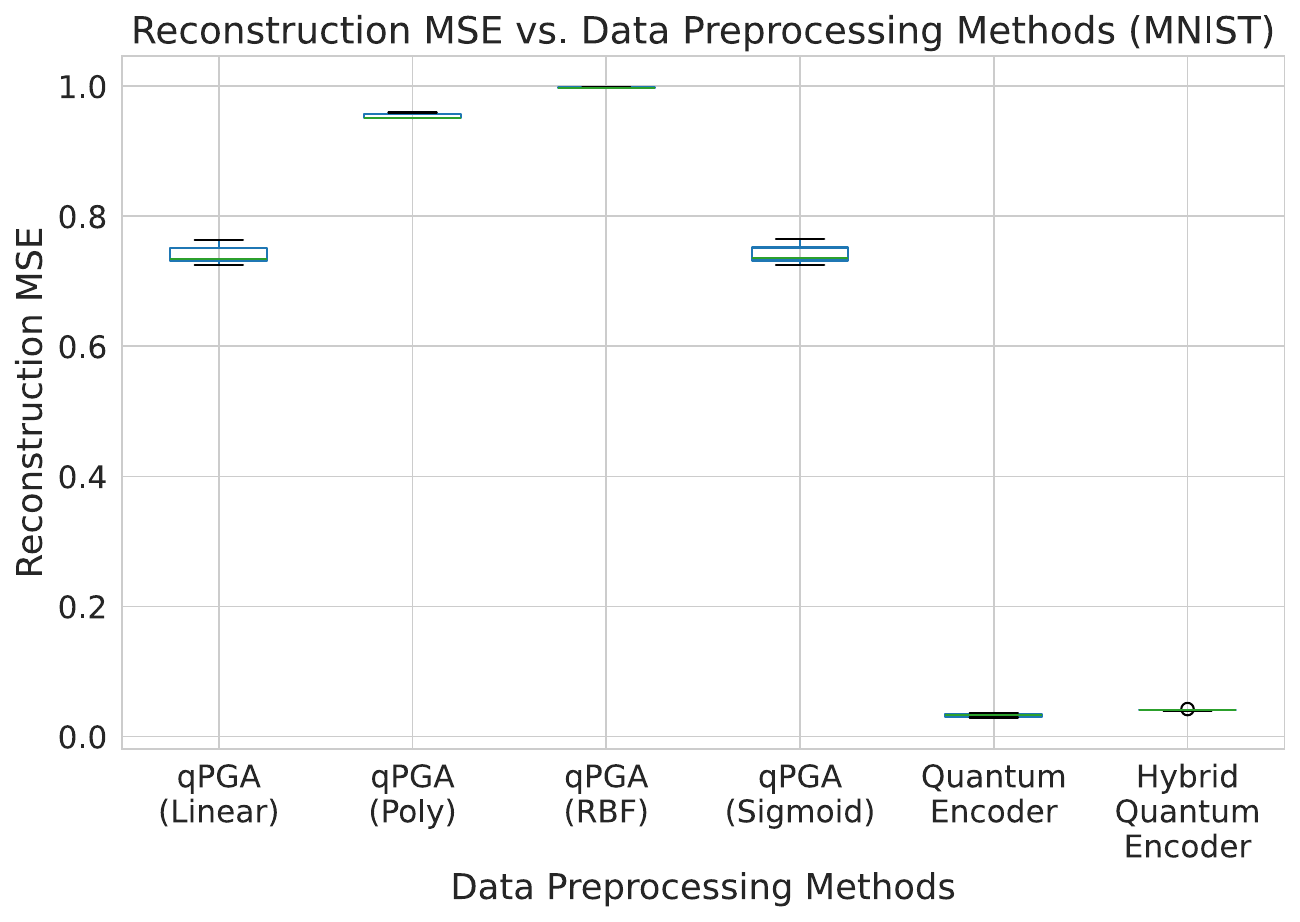}
        \caption{Reconstruction MSE on MNIST}
        \Description{Reconstruction results for MNIST}
        \label{fig:mnist_mse}
    \end{subfigure}
    \medskip
       \begin{subfigure}{0.49\textwidth}
    \centering
        \includegraphics[width=\textwidth]{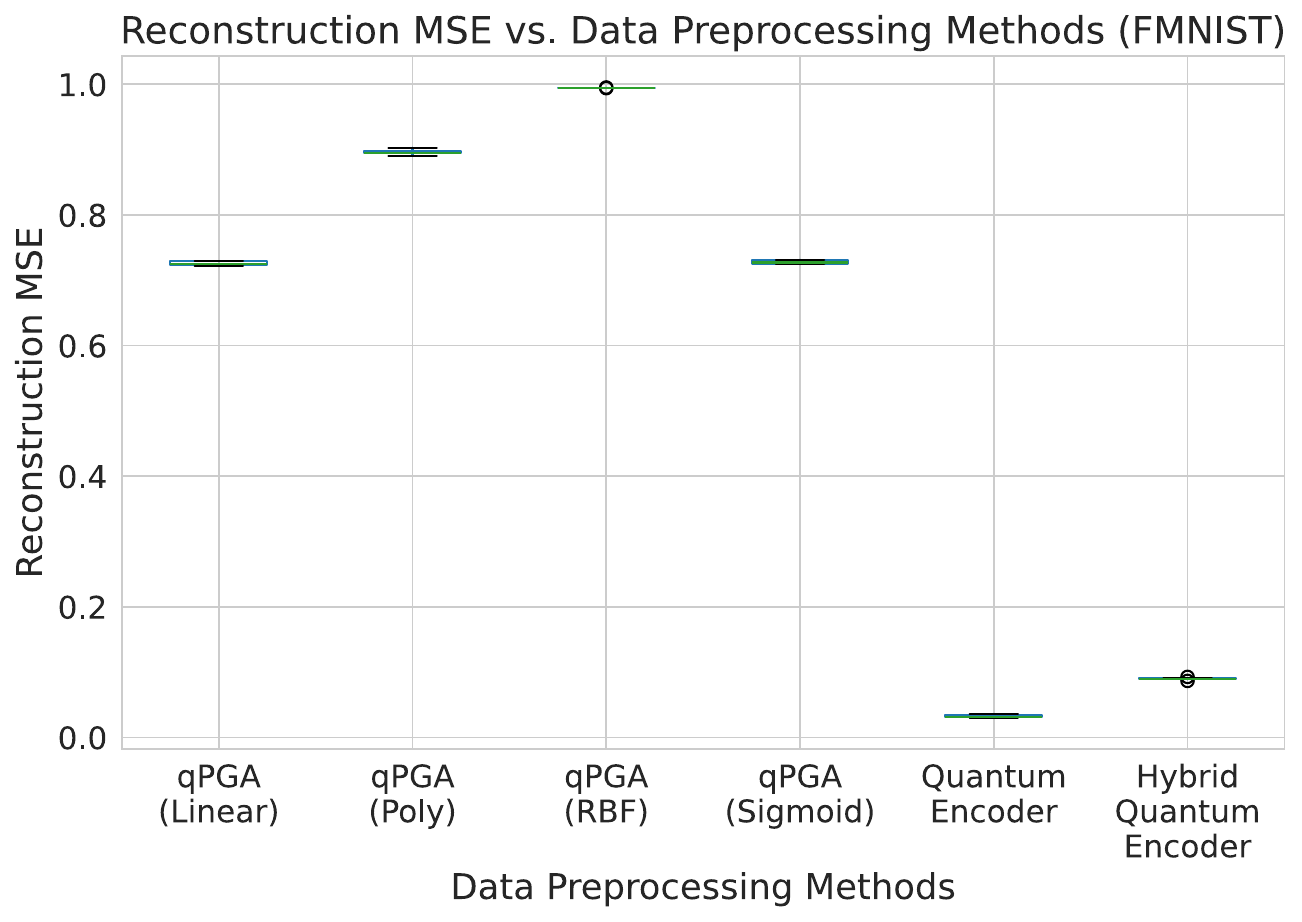}
        \caption{Reconstruction MSE on FMNIST}
        \Description{Reconstruction results for FMNIST}
        \label{fig:fmnist_mse}
    \end{subfigure}
    \medskip
    \begin{subfigure}{\textwidth}
        \centering
        \includegraphics[width=0.49\textwidth]{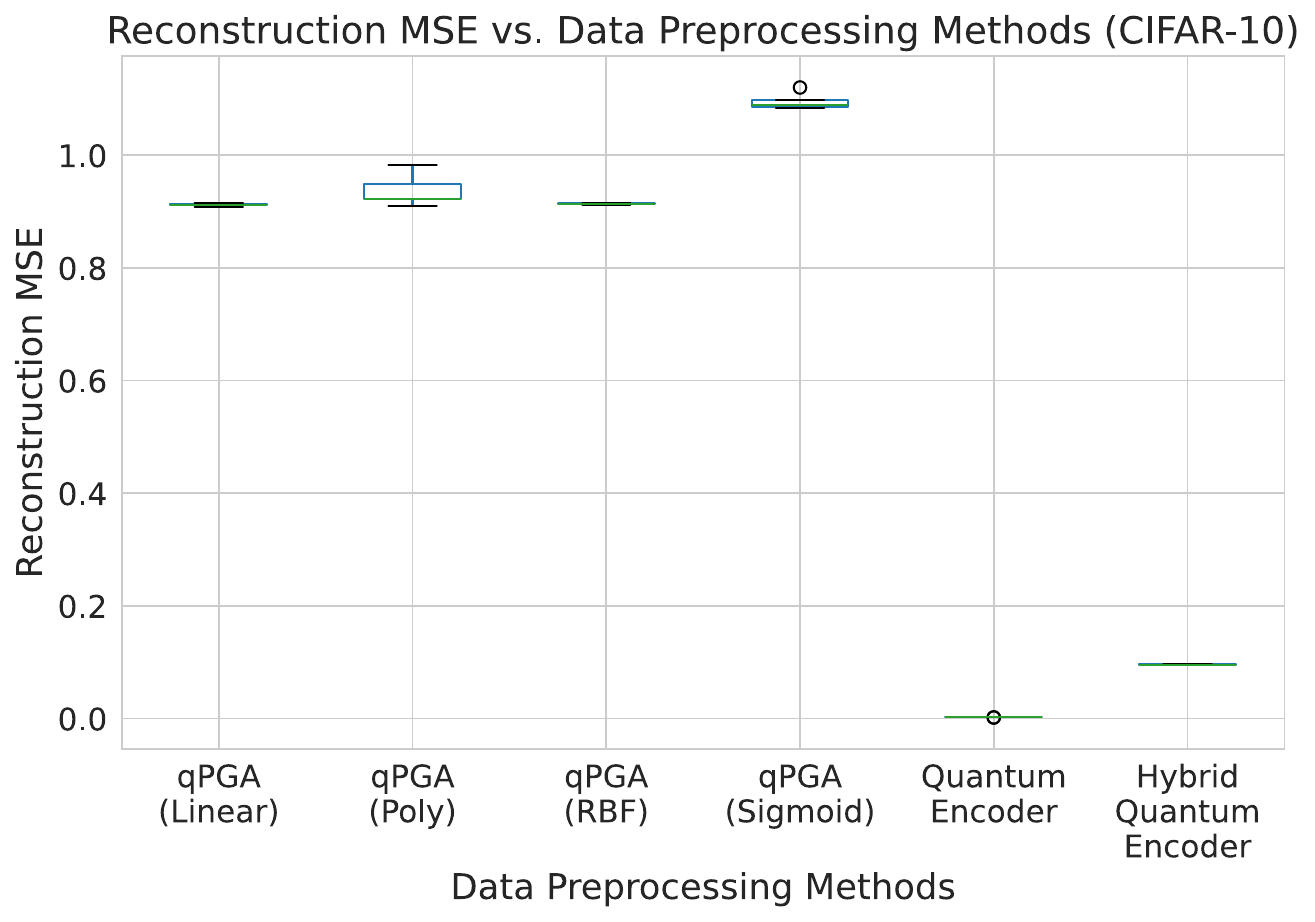}
        \caption{Reconstruction MSE on CIFAR-10}
        \Description{Reconstruction results for CIFAR-10}
        \label{fig:cifar10_mse}
    \end{subfigure}
    \medskip
    \caption{Reconstruction MSE for qPGA (with 4 different kernel feature maps), Quantum Encoder, and Hybrid Quantum Encoder on MNIST, FMNIST, CIFAR-10. Higher MSEs show that reconstruction is more challenging for qPGA compared to the competing methods.}
    \Description{Reconstruction results for 3 datasets}
    \label{fig:mse_comparisons}
\end{figure}
To reconstruct the original data from QE and HQE-reduced latent data, we train their corresponding decoders on the train set and test the trained decoder on each test set. The train and test sets consist of pre-processed versions of the datasets. We repeat this for 5 folds of the dataset and compute the mean and standard deviation. We present these results in box plot format as shown in Fig. \ref{fig:mse_comparisons}.

From the reconstruction MSE results, we note that QE and HQE are invertible methods, i.e., we can reconstruct the original raw data from their latent space representations using the decoder part of their respective autoencoders. 
The same is not possible for the qPGA algorithm. To reconstruct the original data from the latent space representations obtained using the qPGA algorithm, we use the \code{inverse\_transform} command from the \href{https://github.com/geomstats/geomstats}{\code{geomstats}} library, which approximates the inverse mapping from the qPGA-reduced space. We show the difference between the reconstructed and original data using the MSE metric. To compute the MSE metric in the qPGA and QE cases, we employ the geodesic distance as per Eq. (\ref{eq:geodesic_distance}) as the distance metric, as both the kernel feature maps of the original and reconstructed data reside on a UHS. 
For QE, we compute the MSE between the amplitude vectors of the original input quantum states--obtained via amplitude embedding--and those reconstructed from the latent space representations, representing the output quantum states. While fidelity is a common metric for comparing quantum states, we use the MSE based on geodesic distance to ensure a consistent and comparable evaluation framework across all methods considered.

From the reconstruction MSE results shown in Fig. \ref{fig:mse_comparisons}, we deduce that because qPGA is a lossy compression mechanism, it is difficult to reconstruct the original data from their latent space representations using the standard geometric inverse operation. This shows that qPGA as a data pre-processing method is less invertible than QE and HQE. Consequently, from an adversarial attack perspective, our findings provide initial evidence on qPGA's robustness, offering better defence against reconstruction attacks on the latent space representations of high-dimensional data than QE and HQE.
We conclude here that \textit{when employed in the data pre-processing stages, our proposed qPGA algorithm, because it is a lossy compression technique, is less invertible than QE and HQE, demonstrating better resilience in adversarial reconstruction attack scenarios.}

\section{End-to-End Classification Experiments\label{sec:classification}} 
By developing the qPGA algorithm, we have shown that the resulting low-dimensional amplitude vectors effectively capture the structure of high-dimensional datasets in their latent space. 
In this section, we demonstrate the applicability of our proposed qPGA algorithm in downstream QML classification tasks. Specifically, we show how an $N$-dimensional dataset can now be embedded as $D$-dimensional amplitude vectors encoded onto $\lceil\log_2D\rceil$-qubit systems, where $N>D$. These compact quantum representations are subsequently used in binary classification tasks across two distinct types of QML models: a kernel-based approach--the \textit{Quantum Support Vector Machine} \cite{qsvm}, and a variational model--\textit{Quantum Circuit Classifier} \cite{schuld2020circuit}. We describe each of these models below and explain how they are used to classify datasets processed by qPGA, in comparison with those processed by the competing encoder-based methods, QE and HQE.
\subsection{Quantum Support Vector Machine (QSVM)}
The QSVM is the quantum-enhanced version of the classical Support Vector Machine (SVM) used for two-group classification problems in ML \cite{cortes1995support}. The QSVM leverages quantum computing to map classical data to a higher-dimensional quantum feature space using a \textit{quantum kernel}. SVM then operates on this quantum feature space to classify the data points. QSVM has potential advantages over classical SVM in cases where quantum feature spaces can potentially separate data in the feature space more effectively than classical kernels, such as RBF and polynomial kernels. 

The quantum kernel evaluates the similarity between data points by computing the inner products of their corresponding quantum states. This allows for complex, high-dimensional relationships to be captured in a computationally efficient manner. In this section, we show how qPGA enables the use of a small-qubit quantum circuit as the quantum kernel function for QSVM. The quantum circuit consists of an amplitude encoding block to encode the qPGA-processed data into quantum states, followed by their adjoint. These then form the quantum kernel matrix that is employed in SVM to find a decision boundary. 

\paragraph{Quantum Kernel Estimation enhancement by our proposed qPGA algorithm.} The QSVM complexity mainly comes from quantum state preparation and kernel matrix estimation. This step involves encoding classical data into quantum states with a mapping $\phi(x)$ such that the kernel, $K$, is computed as an inner product of quantum states: $K(x,y) = |\langle\phi(x)|\phi(y)\rangle|^2$. The encoding complexity scales as \( \mathcal{O}(\text{poly}(n, \log(d))) \) \cite{rebentrost2014quantum}, offering potential exponential savings compared to classical high-dimensional feature maps which have a complexity of \( \mathcal{O}(n^2 d) \) to \( \mathcal{O}(n^3) \), where \( n \) is the number of samples and \( d \) the number of dimensions. 
Our proposed qPGA algorithm efficiently encodes data into small dimensions embedded as the amplitudes of quantum states of a small number of qubits. Since the computational complexity is poly-logarithmically dependent on dimension $d$, our proposed algorithm demonstrates the potential for further keeping the computational complexity involved in estimating the quantum kernel low. \textit{This advancement underscores the potential of qPGA in quantum kernel estimation, as it provides a scalable and computationally efficient methodology for encoding high-dimensional data into quantum states. By keeping the computational complexity low, qPGA enhances the feasibility of quantum kernel methods, supporting their practical implementation in the fault-tolerant regime of quantum computing.}

\subsubsection{Experiment Setup}
For our experiments, we used 400 data points that were processed using the qPGA algorithm (with 4 normalized kernel feature maps as in Section \ref{sec:exp_results}) and applied 5-fold cross-validation on them for training and testing the QSVM on qPGA-processed datasets. Given $D$ represents the dimension of the latent space, the quantum kernel thus comprises a $\lceil\log_2D\rceil$-qubit quantum circuit with the \code{AmplitudeEmbedding} block, followed by its adjoint. We measure the expectation value of the Hermitian operator at the end of the circuit. We set $D=4$ for MNIST and FMNIST datasets, hence, requiring a 2-qubit quantum circuit, while for the CIFAR-10 dataset, we require a 4-qubit quantum circuit given $D=16$. The choice of these intrinsic dimensionalities, $D$, was explained in Section \ref{sec:exp_results}.

For comparison purposes, we performed the same analysis on the QE-processed data as they are also $D$-dimensional, and lie on the UHS, $\mathcal{S}^{D-1}$. However, for the HQE-processed data, because they are $D$-dimensional lying in the Euclidean space, $\mathbb{R}^D$, we use a $D$-qubit quantum circuit for computing the quantum kernel. The HQE-processed data are phase encoded into the quantum circuit using the \code{AngleEmbedding} block, followed by its adjoint. We again computed the expectation value of the Hermitian operator at the end of circuit evolution. For the HQE-case, when considering the CIFAR-10 dataset, where $D=16$, the kernel matrix has a shape of $2^{16}$ x $2^{16}$,  making it unsimulatable; therefore, we have omitted these results in the result section.

We ran our experiments with the same set of 400 original data points, but processed using three different feature extraction techniques. We used Pennylane's `default-qubit' simulator for all our QSVM classification experiments in this work.
We collected the test accuracies and F1-scores for each fold, and present them in Fig. \ref{fig:qsvm_classification}. 

\subsubsection{Results and Discussions}
\begin{figure}[hbt!]
    \centering
    \includegraphics[width=\textwidth]{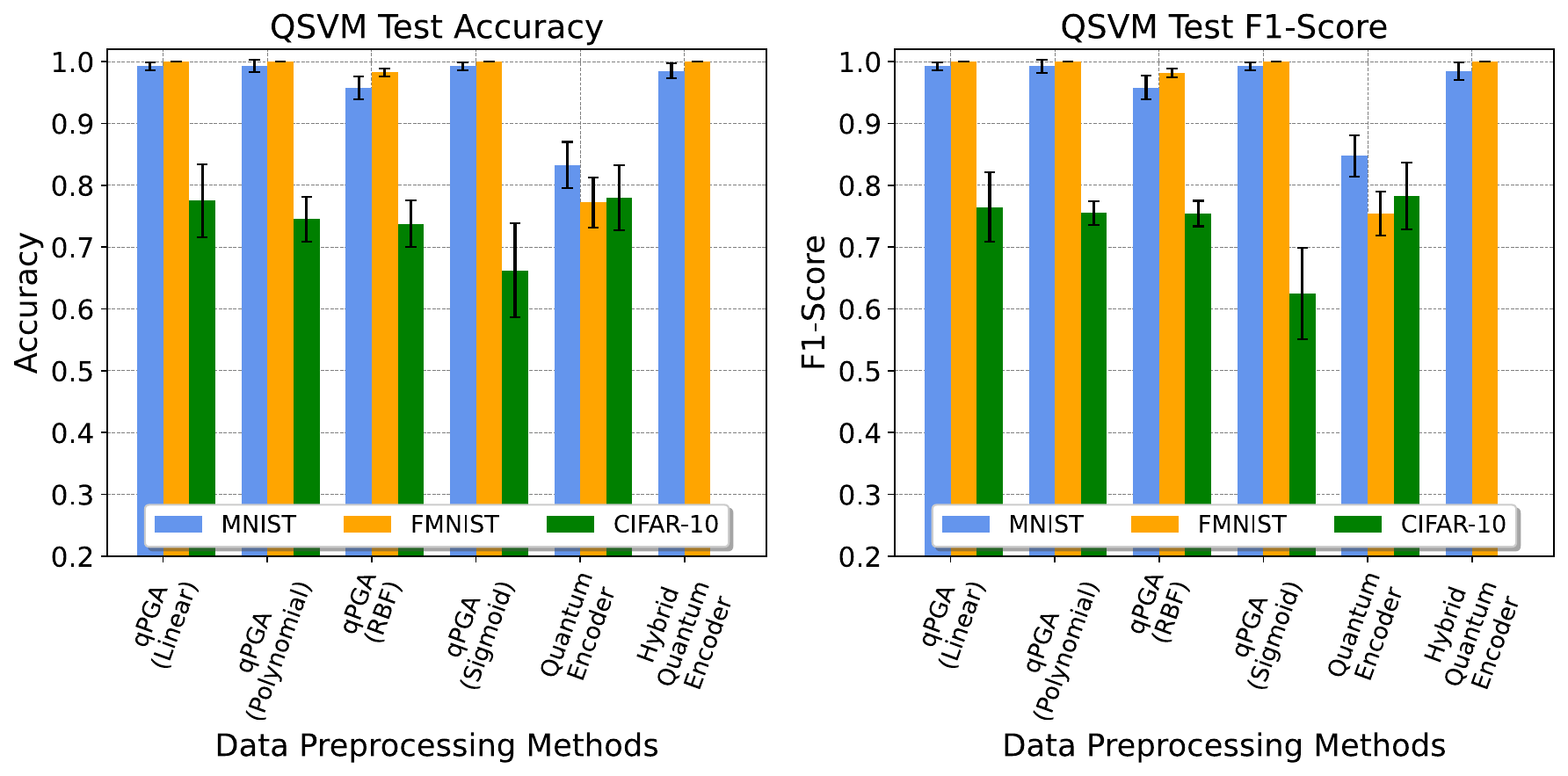}
    \caption{QSVM mean accuracy and F1-score on MNIST (\textcolor{CornflowerBlue}{blue}), FMNIST (\textcolor{YellowOrange}{orange}), and CIFAR-10 (\textcolor{PineGreen}{green}) using qPGA, QE, and HQE. Error bars show variance over 5 folds. qPGA achieves high accuracy and F1-score for MNIST and FMNIST, comparable to HQE. CIFAR-10 results are lower due to dataset complexity; HQE was not tested on CIFAR-10 due to computational limits.}
    \Description{QSVM results Accuracy and F1-score in classification pipeline}
    \label{fig:qsvm_classification}
\end{figure}

Fig. \ref{fig:qsvm_classification} shows the QSVM test accuracies and F1-scores obtained when we classify the data processed using our proposed qPGA (with kernel feature map treated as a tunable hyperparameter, evaluated over linear, polynomial, RBF, and sigmoid kernels), QE, and HQE as competing methods. These experiments were conducted using the \code{default-qubit} simulator from Pennylane. 

The findings for our classification experiments on the MNIST and FMNIST datasets indicate that the qPGA algorithm achieves a classification performance comparable to that of the HQE case, \textit{but with the advantage of needing only a 2-qubit quantum circuit for the quantum kernel computation, as opposed to the HQE scenario's requirement of a 4-qubit device.} Hence, \textit{our proposed qPGA algorithm allows us to match the HQE performance with half the number of qubits.}
Compared to the results on the data processed using QE, which, similar to the qPGA algorithm, requires only a 2-qubit device for kernel computation, the qPGA-processed dataset resulted in much better classification accuracies and F1-scores, and with less variance. This finding is consistent with the trustworthiness and continuity analysis in Fig. \ref{fig:tnc_comparisons}, demonstrating consistency in our proposed qPGA method for efficiently encoding high-dimensional datasets.  
For the CIFAR-10 dataset, we do not obtain a classification performance as high as for the MNIST or FMNIST datasets due to the dataset complexity, but we observe very similar trends across the different methods.

\subsection{Quantum Circuit Classifier\label{vqc}}
The Quantum Circuit Classifier, more commonly referred to as the Variational Quantum Circuit classifier, involves embedding classical data as quantum states in a Hilbert space using a quantum feature map. This is then followed by an \textit{ansatz}, which is the parameterized part of the circuit that involves trainable parameters. At the end of the circuit, we perform measurements followed by classical post-processing to estimate the probability assigned to each class. We use the `parameter-shift' rule, which is a commonly chosen gradient backpropagation method for training the Quantum Circuit Classifier \cite{param_shift}. Next, we describe the circuit model we chose for our Quantum Circuit Classifier, followed by the experiments we conducted.

\subsubsection{Model and Experiments}
\begin{figure}[hbt!]
   \begin{subfigure}{\textwidth}
    \centering
        \includegraphics[width=0.8\textwidth]{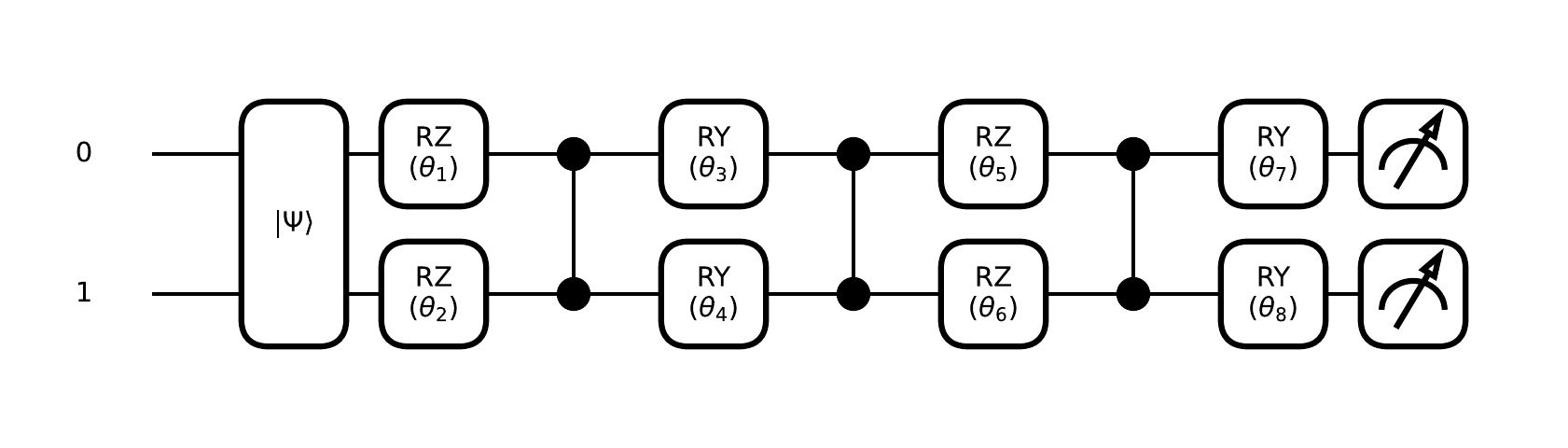}
        \caption{Quantum Circuit Classifier 1}
        \Description{Circuit 1}
        \label{fig:circuit_1}
    \end{subfigure}
    \begin{subfigure}{\textwidth}
     \centering
        \includegraphics[width=\textwidth]{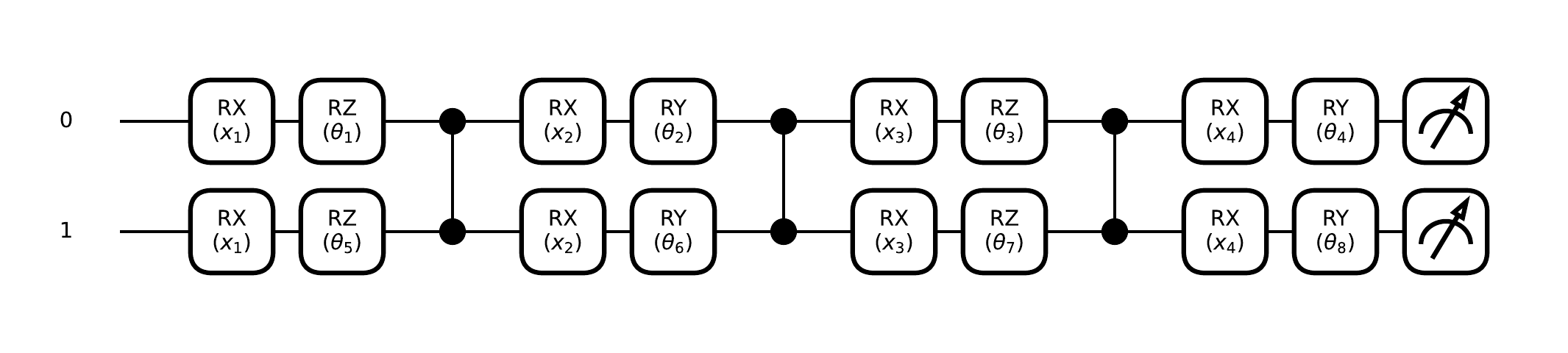}
        \caption{Quantum Circuit Classifier 2}
        \Description{Circuit 2}
        \label{fig:circuit_2}
    \end{subfigure}

    \caption{Quantum circuits for classification with $D=4$ using (a) qPGA and QE, and (b) HQE. For QE, the amplitude embedding block in (a) is omitted after training. We scale the circuits to accommodate higher dimensions, $D$.}
    \Description{Circuits used in this work for Quantum Circuit Classifiers}
    \label{fig:quantum_circuits}
\end{figure}
We considered two quantum circuit configurations based on the space occupied ($\mathbb{R}^D$ or $\mathcal{S}^{D-1}$) by the features extracted using the three feature extraction methods we studied during the data pre-processing stages.
First, since the latent spaces resulting from qPGA and QE consist of normalized $D$-dimensional amplitude vectors that lie on the UHS, $\mathcal{S}^{D-1}$, we encoded these amplitudes onto the quantum states of a $\lceil\log_2D\rceil$-qubit quantum circuit using the \code{amplitude\_embedding} block followed by an \textit{ansatz}. For the QE-processed data, however, we do not require the amplitude embedding block since we concatenate the QE, without performing any prior measurement, with the circuit's \textit{ansatz} directly. This circuit is shown in Fig. \ref{fig:circuit_1}. 

Second, for the HQE-processed dataset, we used phase encoding gates such as RX-gates to encode each feature in the $D$-dimensional feature vector into the quantum circuit. This is because the outputs of the HQE lie in the Euclidean space ($\mathbf{X}\in\mathbb{R}^D$). To keep the number of qubits required in the quantum classifier to $\lceil\log_2D\rceil$, as in the case of the qPGA and QE methods, we adapted the data-loading technique introduced in \cite{hevish_HQSL}, making efficient use of qubits, by alternating the embedding and parameterized gates along each qubit. This qubit-efficient data loading technique allowed us to load multi-dimensional inputs per qubit, at the expense of increased circuit depth. The circuit is as shown as Quantum Circuit Classifier 2 in Fig. \ref{fig:circuit_2}. We note that the \textit{ansatz} designed in Quantum Classifier Circuit 1, as shown in Fig. \ref{fig:circuit_1}, also features a similar arrangement of parameterized gates as in Circuit 2, displayed in Fig. \ref{fig:circuit_2}. The displayed circuits correspond to when $D=4$, i.e., $\mathbf{X}=(\text{x}_1,\text{x}_2,\text{x}_3,\text{x}_4)$. For $D=16$, as in the case of the CIFAR-10 dataset, we require 4-qubit quantum circuits. Hence, we made Circuits 1 and 2 wider, and the CZ-entanglement strategy was such that each pair of qubits was entangled once. 

\noindent
\textit{Measurement and classical post-processing:} We compute the expectation value of the tensor product of Pauli-X operations acting on all the qubits at the end of each circuit and apply the sigmoid function to obtain binary outputs 0 or 1 with a decision threshold of 0.5.

We built the quantum classifiers using Pennylane's \code{default-qubit} device. We trained each of these quantum classifiers on the training set, using the \code{Adam} optimizer with a learning rate of $10^{-2}$, and the binary cross-entropy loss function.
We trained the quantum classifiers for 20 epochs with 5-fold cross-validation. We report the mean test accuracy and F1-score and present the results next. At the end of training, we saved the optimal parameters obtained to build and test our models on multiple devices, as presented in Section \ref{sec:actual_device_results}.

\subsubsection{Results and Discussions}

We report the mean accuracy and F1-score of Quantum Circuit Classifiers on MNIST, FMNIST, and CIFAR-10 datasets in Fig. 9, evaluated over 5 folds. The classifiers were tested using input data pre-processed via our proposed qPGA algorithm and compared against QE and HQE methods. Our findings indicate that, similar to Quantum Support Vector Machine (QSVM) results, the performance of Quantum Circuit Classifiers is highly dependent on the choice of kernel feature map used in the qPGA algorithm. In our experiments, the kernel feature map was treated as a hyperparameter, and we evaluated multiple kernel types, including linear, polynomial, RBF, and sigmoid, to select the most effective mapping for each dataset.
For instance, for the MNIST and FMNIST datasets, the Linear and Sigmoid kernel maps yielded consistently higher accuracies and F1-scores compared to QE and HQE.
The CIFAR-10 dataset, being more complex, generally showed lower performance. 

These observations align with earlier findings (see Fig. \ref{fig:explained_variance_comparisons}), where only 4 principal components from qPGA were sufficient to capture $\beta\geq75\%$ of the dataset variance in MNIST and FMNIST, while 16 principal components were still inadequate for CIFAR-10. This highlights the consistency between achieving high-quality latent representations and achieving strong downstream classification performance.

These results underscore a key advantage of qPGA: \textit{it enables the encoding of high-dimensional classical datasets as low-dimensional amplitude vectors, enabling QML using small-qubit quantum circuit classifiers.
Given the right kernel feature map choice, qPGA yields better classification performance than both QE and HQE, supporting its utility for robust and scalable QML pipelines.}

\begin{figure}[hbt!]
    \centering
    \includegraphics[width=\textwidth]{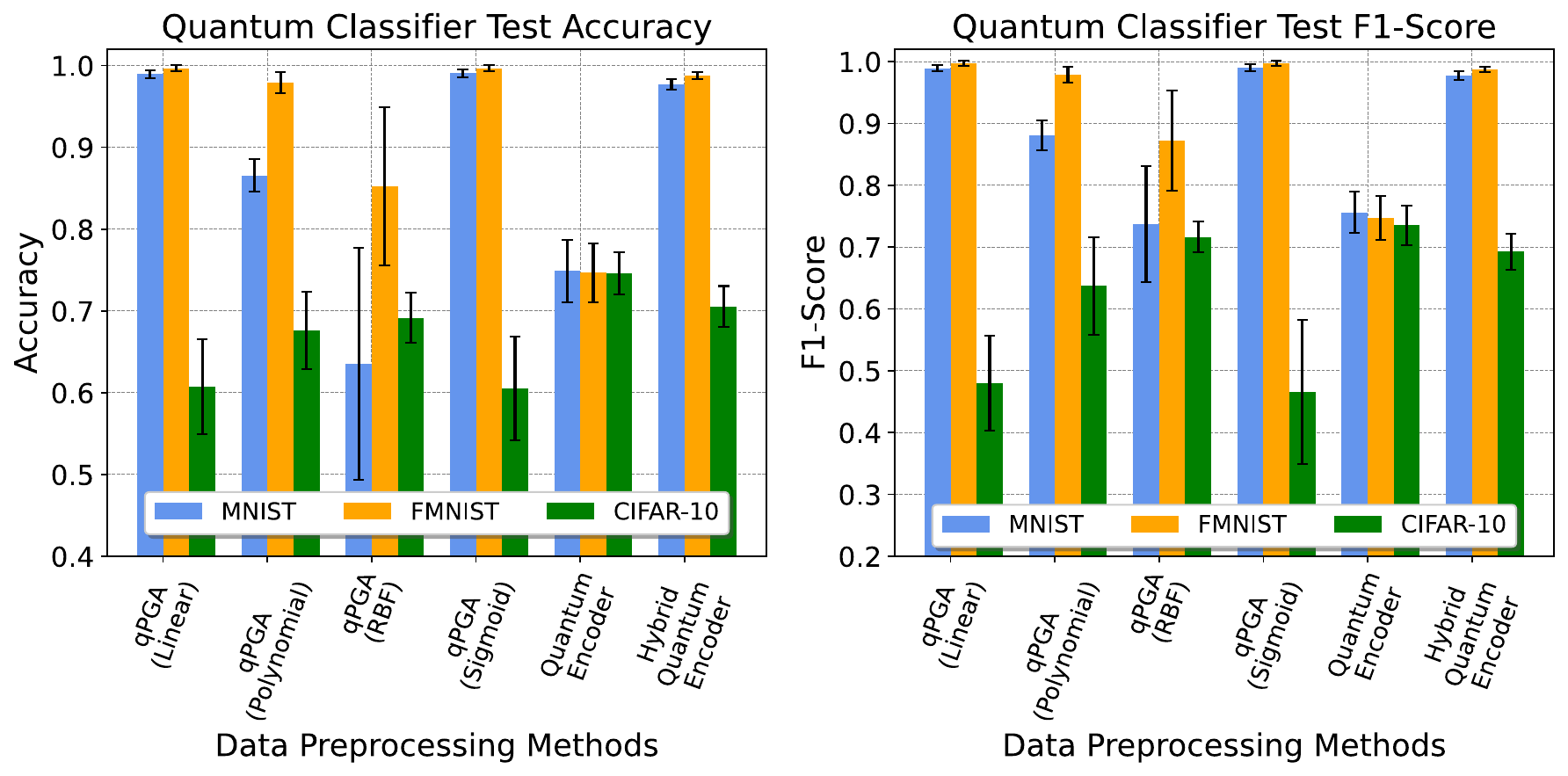}
    \caption{Mean accuracy and F1-score for quantum classifiers on MNIST (\textcolor{CornflowerBlue}{blue}), FMNIST (\textcolor{YellowOrange}{orange}), and CIFAR-10 (\textcolor{PineGreen}{green}) using the three data pre-processing methods. Error bars represent variance over 5 folds. qPGA with linear and sigmoid kernels achieves performance comparable to HQE.}
    \Description{Quantum Circuit Classifier results Accuracy and F1-score in classification pipeline}
    \label{fig:vqc_classification}
\end{figure}

\section{\label{sec:actual_device_results}Evaluations on Actual Hardware and Noisy Quantum Simulators}
To further evaluate the practicality of our proposed qPGA algorithm, we assess its robustness and performance in actual quantum environments. Specifically, we examine how well qPGA-pre-processed data performs on quantum classifiers deployed on actual IBM quantum hardware and noisy simulated quantum devices \cite{qiskit2024}. These experiments are critical to understanding the behavior of our approach under realistic noise conditions and limited qubit availability--common constraints in near-term quantum systems. These experiments provide initial evidence of qPGA's practical potential, particularly its ability to support QML applications using small-qubit systems.
By comparing qPGA against the best-performing competing method (HQE), we aim to demonstrate that qPGA enables more noise-resilient QML pipelines while maintaining strong classification accuracy, even on small-qubit systems. 

\begin{figure}
    \centering
    \includegraphics[width=0.6\linewidth]{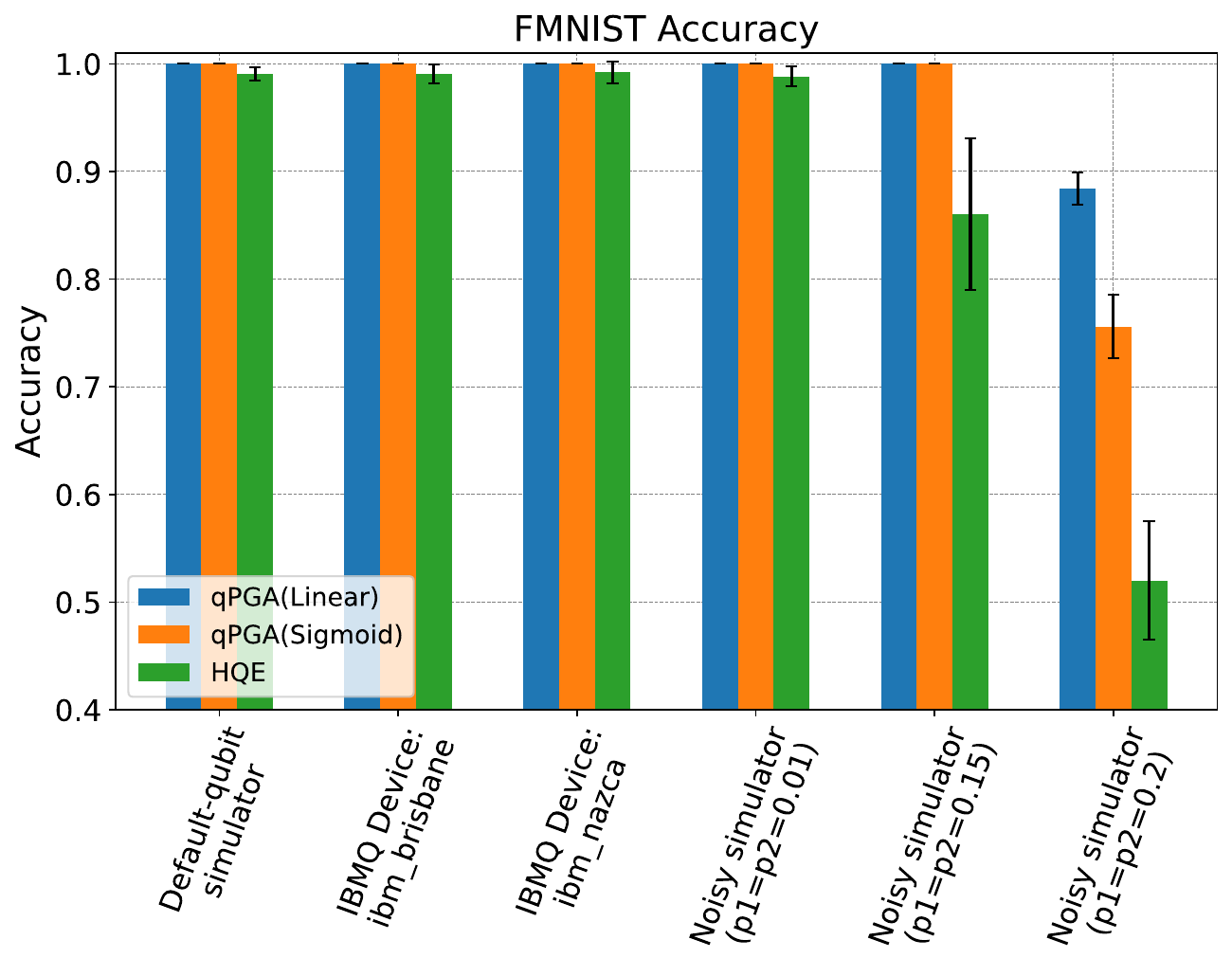}
    \caption{FMNIST test accuracy for qPGA (linear and sigmoid kernel feature maps), and HQE across various devices. 
    qPGA maintains higher performance than HQE for most devices until noise levels reach $p_1=p_2=0.15$; performance degrades significantly at $p_1=p_2=0.2$.}
    \Description{Quantum Circuit Classifier results on multiple devices}
    \label{fig:classification_multiple_device}
\end{figure}

We trained and saved the weights of our Quantum Circuit Classifiers presented in Section \ref{vqc} when run on the `default-qubit' simulator.
We then deployed the Quantum Circuit Classifiers on 2 IBM quantum devices: \textit{ibm\_brisbane, ibm\_nazca} and 3 noisy simulators with simulated noise parameters $p_1$=$p_2$=$p$; where $p=\{0.01, 0.15, 0.2\}$. $p_1$ and $p_2$ represent the depolarizing error due to all single and 2-qubit gates present in our Quantum Circuit Classifiers, respectively. We used the Qiskit platform to access these devices \cite{qiskit2024}. These error parameters characterize the qubit error $P(error) = p$ we described earlier in Section \ref{theory}.
We tested the Quantum Circuit Classifiers on 50 samples of the test set from the FMNIST dataset. 
We carried out tests on the FMNIST dataset processed using (1) qPGA with `Linear' and (2) `Sigmoid' kernel feature maps, and (3) HQE only, as they gave the best performance during noise-free simulations (see Fig. \ref{fig:vqc_classification}). The prediction accuracy results when evaluations are carried out on the different devices are displayed in Fig. \ref{fig:classification_multiple_device}.

Fig. \ref{fig:classification_multiple_device} compares qPGA (with `Linear' and `Sigmoid' kernel feature maps) against HQE on the FMNIST dataset for classification tasks carried out on the noise-free `default-qubit' simulator, actual IBM quantum hardware, and noisy simulators. qPGA method with both `Linear' and `Sigmoid' feature maps maintained high accuracy and robustness (characterized by the height of the error bars) on actual hardware, and also on the noisy simulators up to noise levels $p_1=p_2=0.15$. HQE performed well in noise-free environments and actual hardware but showed significant performance degradation and variability in high-noise scenarios ($p_1=p_2=0.15,0.2$). 
It is important to note that these hardware and noisy simulator experiments were conducted on a limited sample size (N=50 from the FMNIST test set). While these results provide valuable initial indications of qPGA's performance under realistic conditions, more extensive testing with larger sample sizes, across different datasets, and potentially on a wider range of quantum hardware platforms would be necessary to generalize results regarding its noise resilience broadly.

The performance degradation obtained with qPGA methods when tested on the noisy simulator with $p1=p2=0.2$ compared to when $p_1=p2=0.15$ can be explained by Lemma \ref{lemma2} in Section \ref{theory}. As the qubit error rate, $P(error)=p$, increases, we violate the upper bound given the increased noise level in our qubit system, and the representation power of our qPGA algorithm decreases. Thus, the latent space representations do not capture enough dataset variance to yield a high performance.
\textit{From these results, we can conclude that using qPGA in the data pre-processing stages to enable QML with small-qubit systems, we can achieve better noise resilience compared to the HQE method.}

\section{Conclusion \label{sec:conclusion}}
The problem addressed in this paper involves efficiently encoding high-dimensional benchmark datasets onto small-qubit quantum systems, a critical challenge in the context of not only near-term but also fault-tolerant quantum devices to optimize quantum resources and enable scalability of QML algorithms. To tackle this problem, we propose Quantum Principal Geodesic Analysis (qPGA), a novel hybrid encoding technique featuring a classically executed dimensionality-reduction algorithm for feature extraction that leverages the geometric structure of high-dimensional unit-normed classical vectors representing amplitude-encoded quantum states to reduce their dimensions for efficiently encoding them onto small-qubit systems. qPGA ensures that the dataset variance is maximized and the local neighborhood structure of the high-dimensional dataset is preserved in the latent space.

We provided theoretical bounds on the number of qubits required for effective encoding, accounting for the impact of noise in qubit systems under an independent error assumption, to investigate the limits of our algorithm for scalable and practical deployment. Experimental results demonstrate that qPGA outperforms the evaluated quantum encoder-based methods, such as the Quantum Encoder and Hybrid Quantum Encoder, as competing methods in preserving local data structure and, under optimal kernel selection, in downstream classification tasks. Specifically, qPGA preserves the local neighborhood structure of high-dimensional data in the latent space and exhibits preliminary signs of better resistance against reconstruction attacks in adversarial settings compared to the quantum encoder-based methods.

Furthermore, we showed that qPGA integrates seamlessly into end-to-end QML pipelines on noisy simulated and current devices for classification tasks, highlighting its applicability as a hybrid encoding mechanism for both current and future quantum devices. These results underscore the potential of qPGA as a robust and efficient tool for addressing the challenges of encoding high-dimensional classical data onto qubit systems.

Overall, while our results demonstrated the effectiveness and robustness of qPGA in encoding high-dimensional datasets onto small-qubit systems, there remain promising avenues for future research. 
For instance, exploring the theoretical limits of qPGA under more specific and elaborate noise models, particularly those incorporating correlated errors, could provide better insights into how to adapt the algorithm for enhanced robustness and scalability. In addition, while qPGA is specifically designed for amplitude encoding, where input amplitude vectors reside on the UHS, it would be valuable to investigate whether analogous geometry-preserving methods can be developed for other quantum encoding schemes that do not yield unit-normed quantum states with spherical geometry. Furthermore, extensive analyses against a broader range of benchmark methods, beyond Hybrid Quantum Autoencoders and Quantum Autoencoders, would further validate the effectiveness of qPGA, particularly in diverse high-dimensional, real-world data settings. Investigating systematic approaches for optimal kernel selection within qPGA, rather than treating it purely as a hyperparameter, could also further enhance it.

\appendix

\section{Details of Datasets used for our Experiments \label{appendix:datasets}}
In this section, we provide detailed descriptions of the datasets we use in our experiments in this work. We also describe how we pre-process them for our experiments.

\noindent
\textbf{MNIST Dataset:} This dataset is a large collection of 70,000 handwritten digits (0-9) with 10 labels. It contains 60,000 training images and 10,000 test images, each represented as a 28 x 28 grayscale image. We obtain this dataset from \cite{mnist}, and reshape the images to size 8 x 8. In this work, we consider two sets of 600 image samples of labels `0' and `1' respectively for performing binary classification, split and shuffled in a balanced way into 5 folds for cross-fold validation experiments. We process and store images for the MNIST dataset as 1-dimensional vectors of size 64. 

\noindent
\textbf{Fashion-MNIST (FMNIST) Dataset:} This dataset consists of 70,000 grayscale images of fashion items from 10 categories, such as T-shirts, shoes, and bags, intended to serve as a more challenging replacement for the MNIST dataset. Like MNIST, each image is grayscale, consisting of 28 x 28 pixels, with 60,000 images for training and 10,000 for testing. FMNIST can be obtained from \cite{xiao2017fashionmnistnovelimagedataset}. We also reshape these images to size 8 x 8, as in the case of the MNIST dataset. We again consider a two-label case for binary classification, and extract two sets of 600 images corresponding to the labels `0--T-shirt/Top' and 600 images with labels `7--Sneaker', respectively. We split and shuffled the dataset into 5 folds such that each fold consists of an equal distribution of each label. Similar to the MNIST dataset, we flatten the FMNIST images into 1-dimensional vectors of size 64. 

\noindent
\textbf{CIFAR-10 Dataset:} The CIFAR-10 dataset consists of 60,000 color images, each of 32 x 32 pixels, split into 10 classes, with 6,000 images per class \cite{krizhevsky2010cifar}. In this work, we extract images from the `0--Airplane' and `1--Automobile' classes, giving rise to a binary dataset. We only transform the images to grayscale (single-channel) and do not reshape the images to smaller dimensions to prevent loss of information, given the higher complexity of this dataset compared to MNIST and FMNIST. We extract two sets of 600 images from each class, splitting the dataset into 5 folds with an equal distribution of each label for cross-validation experiments. For the CIFAR-10 dataset, we store the image samples as 1-dimensional vectors of size 32x32=1024. 

\section{Quantum Encoder and Hybrid Quantum Encoder as Competing Methods\label{appendix0:methods}}

In this section, we provide detailed descriptions of two established quantum-dependent methods (Quantum Encoder and Hybrid Quantum Encoder) to benchmark our qPGA algorithm in the data pre-processing stage for encoding high-dimensional datasets into small-qubit systems. Specifically, qPGA and the competing methods reduce the dimensions of classical data ($\mathbf{X} \in \mathbb{R}^N$) to encode them as low-dimensional data onto qubit systems. The qPGA algorithm and Quantum Encoder (QE) compress the dimension from $\mathbb{R}^N$ to $\mathcal{S}^{D-1}$, where $D<N$ represents the dimension of the space spanned by the amplitude vectors of the quantum states to be encoded using $\lceil\log_2D\rceil$ qubits. For the Hybrid Quantum Encoder (HQE), the outputs lie in the $\mathbb{R}^D$ space due to the projective measurements applied at the end of a $D$-qubit quantum circuit in the encoder part. We describe these methods and how we adapt them for benchmarking our proposed qPGA algorithm next. 

\subsection{Competing Method 1: Quantum Autoencoder \texorpdfstring{$(\mathbb{R}^N \mapsto \mathcal{S}^{D-1})$}{Quantum Autoencoder} \label{subsec:method2}}

The quantum autoencoder, similar to classical autoencoders, learns low-dimensional representations of data in a higher-dimensional space. We adopt the Quantum Autoencoder structure devised by \cite{romero_quantum_2017} in this work. 

\begin{figure}
    \centering
    \includegraphics[width=\linewidth]{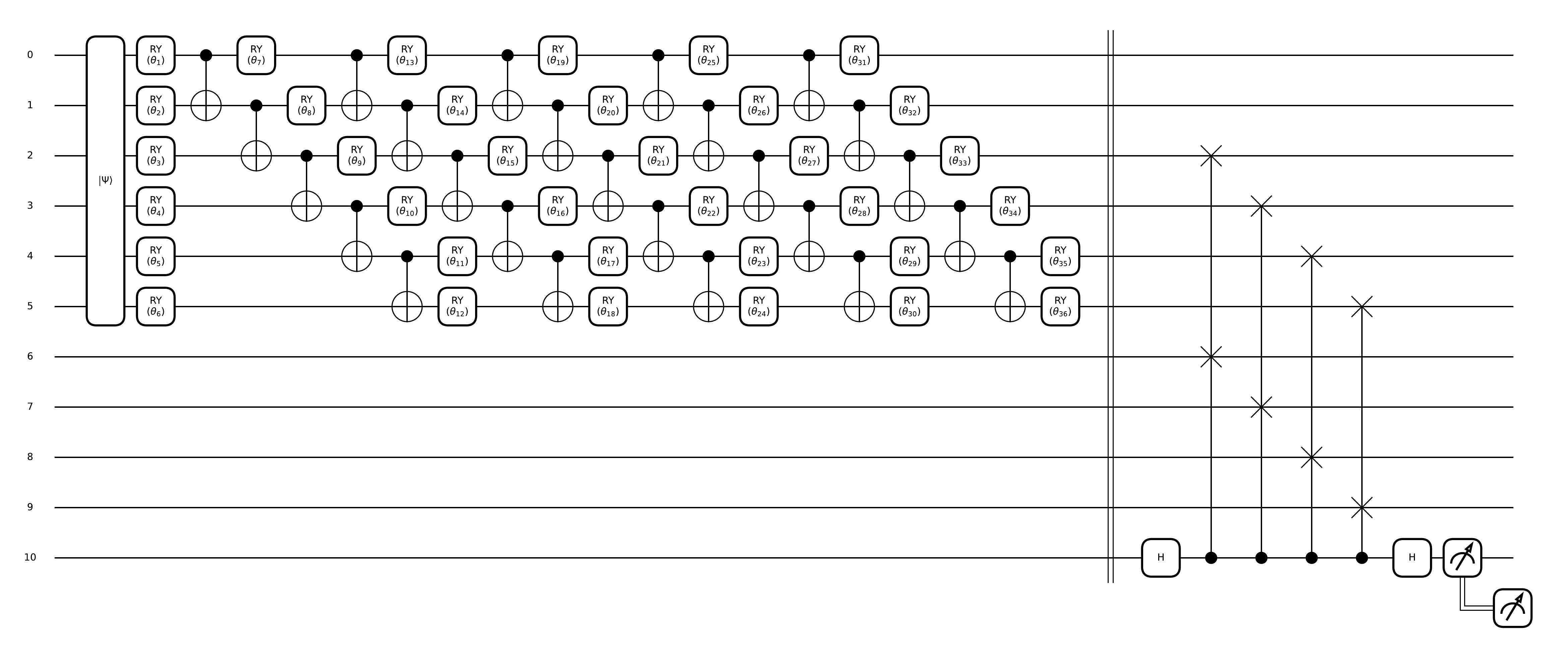}
    \caption{Quantum autoencoder adapted from \cite{romero_quantum_2017} to reduce $N=64$ dimensions to $D=4$. $\lceil\log_2N\rceil = 6$ input qubits (0-5) are mapped to $\lceil\log_2D\rceil = 2$ latent qubits (0-1), with SWAP gates between the trash (2-5) and reference qubits (6-9). Measurement is applied to qubit 10 to compute fidelity during training. 
    }
    \Description{Circuit used for Quantum Autoencoder}
    \label{fig:quantum_encoder}
\end{figure}

Classical data, $\mathbf{X}\in\mathbb{R}^N$, is encoded into quantum states via amplitude embedding, where the entries of $\mathbf{X}$ are normalized to satisfy the unit $\ell_2$-norm constraint of quantum states. To align the dimension of the classical vector with the Hilbert space dimension ($2^n$), zero-padding is applied if $N$ is not a power of two, where $n=\lceil\log_2N\rceil$ denotes the number of input qubits.
The input state, represented by $n=n_{trash}+n_{latent}$ qubits, consists of the trash ($n_{trash}$ qubits) and latent ($n_{latent}$ qubits) space. The trash space consists of redundant information that can be discarded, while the latent space captures maximally compressed information from the input space \cite{ma_compression_2023}. We also include zero-initialized reference states, represented by $n_{ref}$ $\ket{0}$ qubits, such that, $n_{ref} = n_{trash}$. By performing the SWAP routine between the trash and reference states, we can extract the optimally compressed latent space. We also include an auxiliary qubit and a classical register for storing the measurement of the auxiliary qubit to determine the overlap of the trash and reference states of the system, maximizing the fidelity between the input and latent states during training. 
Inspired by the work in \cite{romero_quantum_2017}, we use the \code{RealAmplitudes} \textit{ansatz}, with 5 repetitions as the parameterized part of the quantum circuit in the quantum autoencoder. This \textit{ansatz} is a two-local parametrized circuit consisting of alternating rotation (RY) and CX-entanglement layers. Moreover, it is suitable for benchmarking qPGA, as the prepared quantum states in QE also consist of only real amplitudes. This ensures that we capture maximal information in just the real amplitudes of the quantum states represented by the latent qubits. The amplitude vectors here also lie in the UHS, $\mathcal{S}^{D-1}$, where $D=2^{n_{latent}}$, similar to the latent space of the outputs of our proposed qPGA algorithm as explained in Section \ref{qPGA_alg}. We display an example of the quantum autoencoder circuit we use in this work in Fig. \ref{fig:quantum_encoder}.

We describe the fidelity loss function, $\mathcal{L}$ as, $\mathcal{L} = 1-\frac{2}{M}L$, where $M$ is the number of shots and $L$ is the number of times the auxiliary qubit is in the $\ket{1}$ state, showing maximal overlap between the trash and reference states. Hence, we aim to maximize the function $\mathcal{L}$, i.e., minimize the ratio $\frac{L}{M}$. This represents the cost function we use while training the Quantum Autoencoder. In Section \ref{sec:classification}, we demonstrate how we can use the trained quantum encoder (QE) to compress the dimension of classical data into its low-dimensional latent space representation represented by the amplitudes of the quantum states of latent qubits. These low-dimensional quantum states are then fed directly into quantum systems represented by quantum circuits involved in QML tasks.
 
\subsection{Competing Method 2: Hybrid Quantum Autoencoder \texorpdfstring{$(\mathbb{R}^N \mapsto \mathbb{R}^D)$}{Hybrid Quantum Autoencoder} \label{subsec:method3}}
\begin{figure}[ht]
    \centering
    
    \begin{subfigure}[b]{\linewidth}
        \centering
        \includegraphics[width=\linewidth]{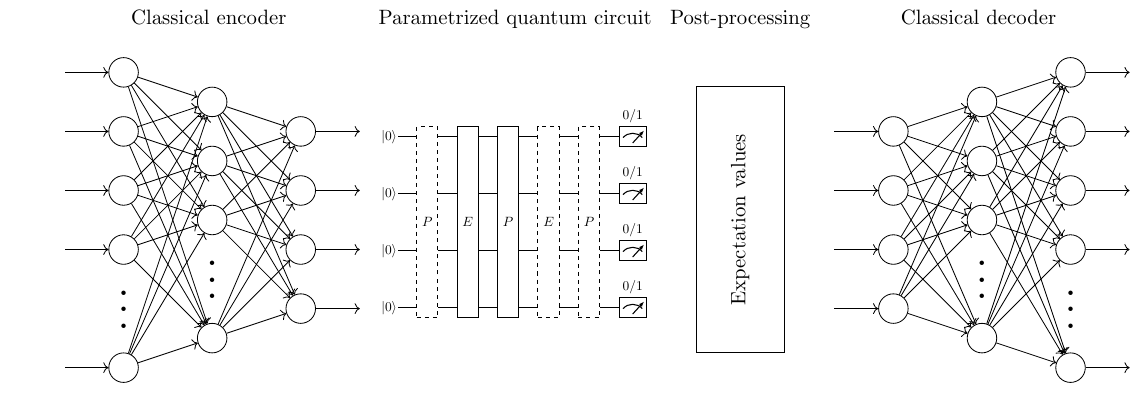}
        \caption{Hybrid Quantum Autoencoder Architecture}
        \Description{General Architecture for Hybrid Quantum Autoencoder}
        \label{fig:hqe_architecture}
    \end{subfigure}
    \hfill
    \begin{subfigure}[b]{0.9\linewidth}
        \centering
        \includegraphics[width=\linewidth]{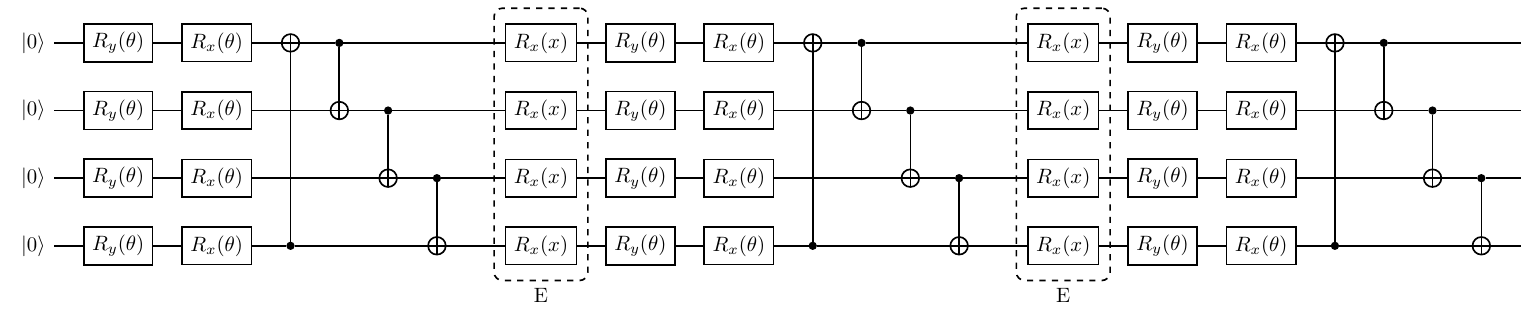}
        \caption{Parametrized Quantum Circuit in the bottleneck layer of the Hybrid Quantum Autoencoder from \cite{sakhnenko_hybrid_2022}}
        \Description{Bottleneck layer}
        \label{fig:circuit_hqe}
    \end{subfigure}
    
    \caption{Hybrid Quantum Autoencoder adapted from \cite{sakhnenko_hybrid_2022}. (a) Architecture with $N$ input/output nodes, and $D$ classical nodes at the bottleneck layer. (b) Parametrized/Variational Quantum Circuit in the bottleneck layer; the expectation values at post-processing are used as classical features for the classical decoder.}
    \Description{Hybrid Quantum Autoencoder used in this work}
    \label{fig:hqe_combined}
\end{figure}

The hybrid quantum autoencoder (HAE) architecture is another example of a quantum computing-based autoencoder. It consists of a classical encoder, a parameterized quantum circuit (or VQC) at the bottleneck layer, and a classical decoder, to reconstruct the input data from the outputs of the VQC. We adopt the HAE structure proposed in \cite{sakhnenko_hybrid_2022} to reduce the dimension of data from N to D. Fig. \ref{fig:hqe_architecture} shows the general architecture of HAE that we use in this work.

The encoder part of the autoencoder (HQE) consists of classical neural network layers followed by a quantum layer consisting of a quantum circuit. We use the \code{Tanh} activation after each classical layer. The classical layers map $N$-dimensional classical input down to $D$-dimensional features that are encoded into the quantum states of a $D$-qubit quantum circuit with phase encoding. 
The quantum circuit lies at the bottleneck portion of the autoencoder and further processes the latent space without any further reduction in dimension. As explained by Sakhnenko et al. \cite{sakhnenko_hybrid_2022}, the quantum circuit improves the latent representation of the high-dimensional data compared to a classical counterpart of the quantum circuit. For the bottleneck layer, we employ the quantum circuit used in \cite{sakhnenko_hybrid_2022}, given it was the best circuit design (out of 32 tested) exhibiting the best improvement in performance. This quantum circuit is shown in Fig. \ref{fig:circuit_hqe}. The post-processing layer converts the output of the quantum circuit into classical features ($\mathbf{X}_{latent}\in\mathbb{R}^D$) which are then fed into $D$-input nodes of the classical decoder. The decoder part consists of classical layers, with output dimensions of $N$.  
The HAE can then be trained in an end-to-end manner using classical optimizers, e.g., \code{Adam}, to minimize a loss function that compares the similarity between the input and reconstructed data. 

After training, we extract the hybrid encoder to compress the original data to the latent space where $\mathbf{X}_{latent} \in \mathbb{R}^D$, contrary to the outputs of the Quantum Encoder and qPGA methods, where $\mathbf{X}_{latent} \in \mathcal{S}^{D-1}$. 
\section{Details of Metrics used to Assess qPGA against Existing Methods \label{appendix:metrics}}
In this section, we describe in detail the metrics that are commonly used to assess the effectiveness of different dimension reduction methods. The metrics are \textit{Explained Variance}, \textit{Co-ranking matrix}, \textit{Trustworthiness}, and \textit{Continuity}. We use the Explained Variance metrics to evaluate the performance of our qPGA algorithm in capturing a specified proportion $\beta$ of the original dataset's variance in its latent space representations. The remaining metrics determine how well the local structure of the high-dimensional original dataset is maintained in its low-dimensional representations. These metrics hence ensure that the latent space vectors capture these characteristics of the high-dimensional dataset, ensuring its effective and accurate embedding onto small-qubit systems. We detail these metrics next:  
We evaluate the effectiveness of our dimension reduction techniques by using the following metrics. 

\subsection{Explained Variance}
In PCA, explained variance indicates how much of the dataset's variance is retained after transforming it to a lower-dimensional space. Similarly, in this context, the explained variance measures how effectively the qPGA algorithm captures the variance of the original high-dimensional dataset within each of the principal geodesic components in its latent space representation on the UHS, denoted as $\mathcal{S}^{D-1}$, where $D$ is the dimension of the latent space. The explained variance for each principal component is indicated by the singular values. We compute the cumulative explained variance, showing how much of the dataset variance is captured by the first $D$ principal components. This metric thus serves as an indicator of how well the qPGA algorithm captures the dataset variance after dimensionality reduction.

\subsection{Co-ranking Matrix}
The co-ranking matrix serves as a method to assess the performance of dimensionality reduction methods \cite{lee_quality_2009, lueks2011evaluatedimensionalityreduction} by considering the neighborhood relationships between high and low-dimensional spaces. The co-ranking matrix is particularly useful for assessing how well the neighborhood structure is retained after dimensionality reduction, with an emphasis on maintaining the relative ordering of pairwise distances or similarities among points. In the original high-dimensional space, each data point, denoted as $\mu$, has neighbors that are ranked based on their proximity to $\mu$. After dimensionality reduction, the data point $\mu$ will possess neighbors in the reduced-dimensional space, also ranked by their distances. The co-ranking matrix compares these rankings (in high-dimensional versus low-dimensional spaces) for every point pair in the dataset, verifying if points closely ranked in the original space maintain a similar ranking in the reduced space. We formally describe the co-ranking matrix structure below.

\noindent
\paragraph{Definition.} Given a dataset with $n$ points, the co-ranking matrix $Q$ is an $n$ x $n$ matrix where both the rows and columns represent the ranks of the points in the original high-dimensional space and the latent space, respectively. Each element $Q(i,j)$ of the matrix counts how often pairs of points that were in the $i^{th}$ rank in the high-dimensional space have been moved to the $j^{th}$ rank in the latent space. The diagonal entries $Q(i,i)$ show how many pairs of points retained their exact rank after dimensionality reduction, while the off-diagonal entries show how much the rankings of the points changed after dimensionality reduction. 

Hence, ideally, a good dimensionality reduction technique will have most of the entries lying on the main diagonal in the co-ranking matrix, indicating that the ranks are preserved. The off-diagonal entries indicate mismatches in the neighborhood structure between the high-dimensional and latent spaces. 

\subsection{Trustworthiness and Continuity}
These metrics, similar to the co-ranking matrix, measure how well the neighborhood structure of data is preserved after dimension reduction \cite{stasis_semantically_2016}. \textit{Trustworthiness}, $T$, measures how well the low-dimensional embeddings preserve the local neighborhood structure of the high-dimensional data. Specifically, it assesses the degree to which the nearest neighbors in the low-dimensional space are also nearest neighbors in the high-dimensional space. The Euclidean distance of point $i$ in high-dimensional space is measured against its $k$ closest neighbors using rank order, and the extent to which each rank changes in low-dimensional space is measured. 
On the other hand, \textit{continuity}, $C$, measures the inverse property: how well the neighborhood structure in the high-dimensional space is preserved in the low-dimensional embedding. It ensures that points that are close in the high-dimensional space remain close in the low-dimensional representation. 
Given the following notations: 
\begin{itemize}
    \item \( n \) is the number of points,
    \item \( k \) is the number of neighbors considered,
    \item \( U_i(k) \) is the set of points that are among the \( k \)-nearest neighbors of point \( i \) in the high-dimensional space but not in the low-dimensional space,
    \item \( \text{rank}(j) \) is the rank of point \( j \) in the sorted list of distances from point \( i \) in the \textit{low-dimensional }space,
    
    \item \( V_i(k) \) is the set of points that are among the \( k \)-nearest neighbors of point \( i \) in the low-dimensional space but not in the high-dimensional space,
    \item \( \text{rank}(m) \) is the rank of point \( m \) in the sorted list of distances from point \( i \) in the \textit{high-dimensional} space,
\end{itemize}
we can express these metrics as follows: 
\begin{equation}
\text{Trustworthiness}, T(k) = 1 - \frac{2}{n k (2n - 3k - 1)} \sum_{i=1}^{n} \sum_{j \in U_{i}(k)} (\text{rank}(j) - k)
\end{equation}
\begin{equation}
\text{Continuity}, C(k) = 1 - \frac{2}{n k (2n - 3k - 1)} \sum_{i=1}^{n} \sum_{m \in V_{i}(k)} (\text{rank}(m) - k)
\end{equation}
As explained in Section \ref{subsec:metrics}, for points lying on the UHS, we use the geodesic distance as the metric for computing the distance-based rankings of each $k$ neighbor to point $i$ to obtain the sets $U_i$ and $V_i$. 

\begin{figure}[ht]   
    \centering
    \begin{subfigure}{0.49\textwidth}
        \includegraphics[width=\textwidth]{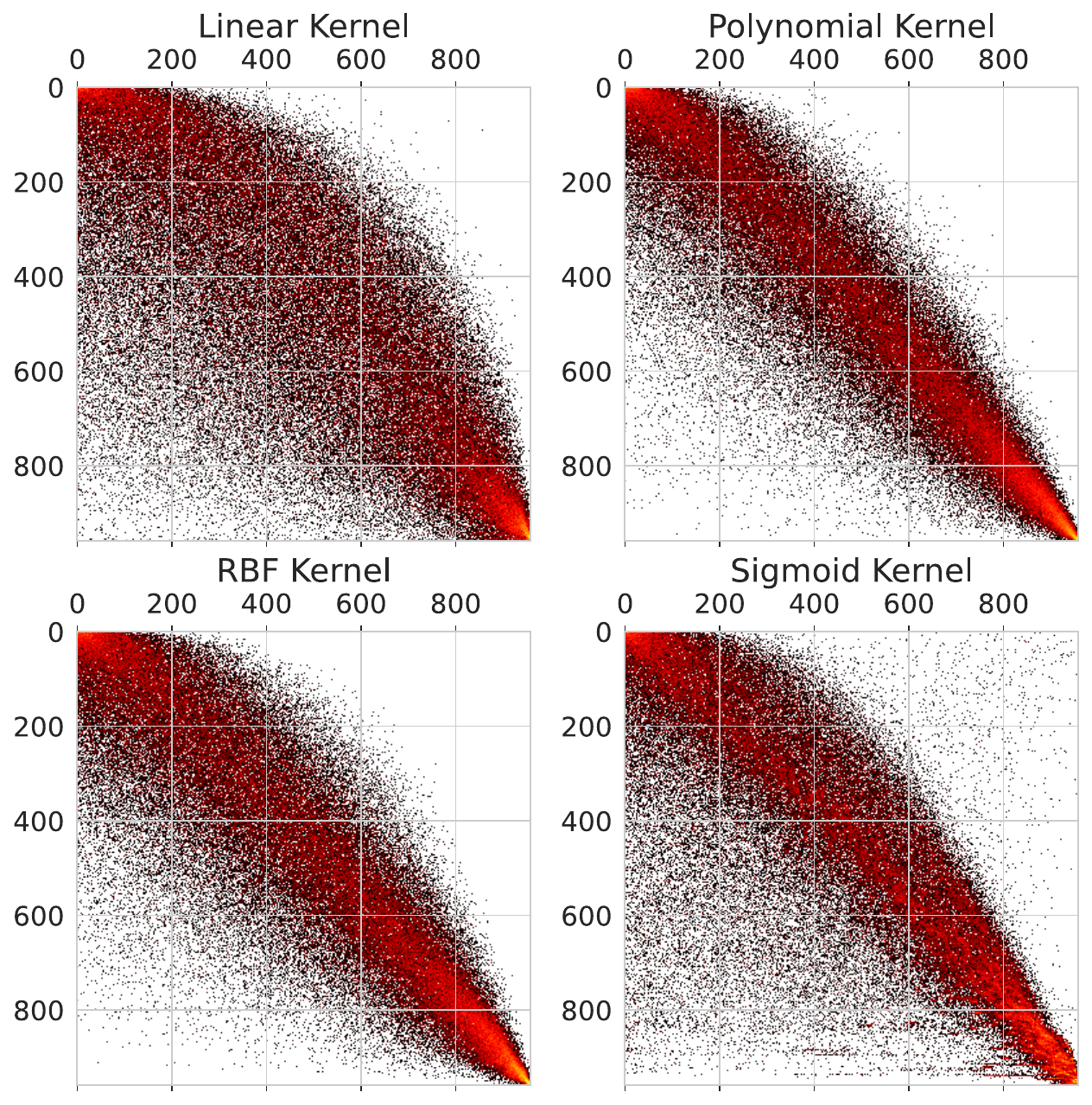}
        \caption{$D=4$ dimensions}
        \Description{Coranking for D=4}
        \label{fig:cifar_4_dim_corank}
    \end{subfigure}
    \begin{subfigure}{0.49\textwidth}
        \includegraphics[width=\textwidth]{results/fig8.pdf}
        \caption{$D=16$ dimensions}
        \Description{Coranking for D=16}
        \label{fig:cifar_16_dim_corank}
    \end{subfigure}
    \begin{subfigure}{\textwidth}
        \centering
        \includegraphics[width=0.49\textwidth]{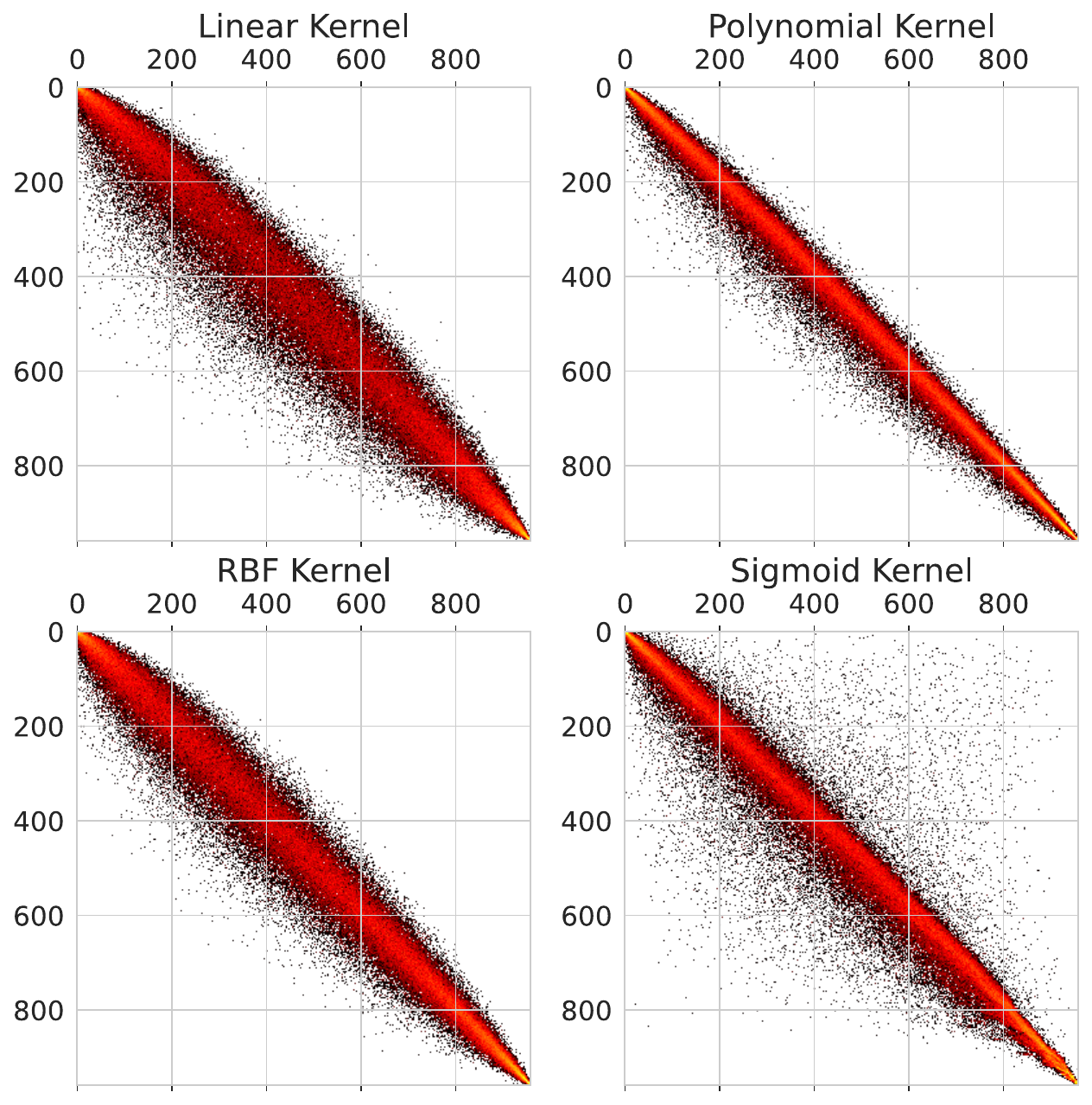}
        \caption{$D=32$ dimensions}
        \Description{Coranking for D=32}
        \label{fig:cifar_32_dim_corank}
    \end{subfigure}
    \caption{Co-ranking matrices on CIFAR-10 at latent dimensions, $D = 4, 16, 32$. Increasing $D$ improves the rank preservation between high-dimensional data and their low-dimensional representations.}
    \Description{Coranking results for increasing D}
    \label{fig:cifar_dims_corank}
\end{figure}
\section{Effect of Increasing Number of Principal Components on Quality of Latent Space Representations\label{appendix1} from qPGA Algorithm}
In this study, we illustrate how the qPGA algorithm enhances the latent space representation of high-dimensional data by extracting an increasing number of principal components. Capturing more principal components leads to a greater percentage of the dataset variance being represented in the latent space. Fig. \ref{fig:cifar_dims_corank} demonstrates that as the number of intrinsic dimensions (or the number of principal components from our proposed qPGA algorithm) $D$ increases from 4 to 16 and 32, the clustering distribution becomes more concentrated along the diagonal. In this work, we consider $D=16$ for the CIFAR-10 dataset as this represents a reasonable portion of the dataset variance being retained in the latent space representation while keeping the number of qubits required, hence, the amount of noise, to represent them low enough in a subsequent quantum system.

This shows that the local neighborhood structure (or rank relationships) of the high-dimensional data is better preserved in their latent space representations when $D$ and the proportion of dataset variance captured increase.

\bibliographystyle{ACM-Reference-Format}
\bibliography{paper_refs, references}

\end{document}